\documentclass[a4paper,11pt]{article}

\usepackage[margin=1in]{geometry}

\usepackage{enumitem}
\usepackage{graphicx}
\usepackage{caption,subcaption}
\usepackage{authblk}
\graphicspath{{Figures/}}

\usepackage{url}

\usepackage[numbers,sort&compress]{natbib}

\usepackage{framed}

\usepackage{tikz}
\usetikzlibrary{fit}
\usetikzlibrary{intersections}
\tikzstyle{vertex}=[circle,fill,inner sep=1pt]
\tikzstyle{point}=[circle,fill,inner sep=1pt]
\tikzstyle{spoint}=[circle,fill,inner sep=0.88pt]
\usetikzlibrary{calc}

\usepackage{amsthm}
\usepackage{amsmath} 
\usepackage{amssymb}

\usepackage{thmtools}
\usepackage{thm-restate}

\usepackage{xspace}
\usepackage{rotating}
\usepackage[color=green!20]{todonotes}

\newtheorem{theorem}{Theorem}
\newtheorem*{claim}{Claim}
\newtheorem{lemma}[theorem]{Lemma}
\newtheorem{lem}[theorem]{Lemma}
\theoremstyle{definition}
\newtheorem{ass}{Assumption}
\newtheorem{example}[theorem]{Example}
\newtheorem{definition}[theorem]{Definition}

\usepackage{comment}

\includecomment{soda}
\excludecomment{full}

\def\pgag{\textsc{Point Guard Art Gallery}\xspace}

\def\pg{\textsc{Point Guard}\xspace}

\def\ray{\ensuremath{\text{ray}}}
\def\seg{\ensuremath{\text{seg}}}
\def\disk{\ensuremath{\text{disk}}}
\def\dist{\ensuremath{\text{dist}}}

\def\vis{\ensuremath{\text{Vis}}}
\def\P{\ensuremath{\mathcal{P}}\xspace }
\def\cone{\ensuremath{\textrm{cone}}\xspace }
\def\apex{\ensuremath{\textrm{apex}}\xspace }

\def\grid{\ensuremath{\alpha \textrm{-grid}}\xspace }

\def\gridconstant{\ensuremath{11}\xspace}

\def\R{\ensuremath{\mathbb{R}}}
\def\N{\ensuremath{\mathbb{N}}}

\def\round{\ensuremath{\textrm{round}}\xspace }
\def\diam{\ensuremath{\textrm{diam}}\xspace }

\usepackage{etoolbox}
\usepackage{aliascnt}

\newcounter{Bew1}
\newcounter{Bew2}
\newcounter{Def1}

\renewcommand{\leq}{\leqslant}
\renewcommand{\geq}{\geqslant}

\title{An Approximation Algorithm for the Art Gallery Problem\footnote{supported by the ERC grant PARAMTIGHT: "Parameterized complexity and the search for tight complexity results", no. 280152.}}

\author[1]{\'{E}douard Bonnet\thanks{edouard.bonnet@lamsade.dauphine.fr}}
\author[1]{Tillmann Miltzow\thanks{t.miltzow@gmail.com}}
\affil[1]{Institute for Computer Science and Control
Hungarian Academy of Sciences\newline(MTA SZTAKI)}
\date{}

\begin{document}

\maketitle

\begin{abstract}
  Given a simple polygon $\mathcal{P}$ on $n$ vertices, two points $x,y$ in $\mathcal{P}$ are said to be visible to each other if the line segment between $x$ and $y$ is contained in $\mathcal{P}$. 
  The \pgag problem asks for a minimum set $S$ such that every point in $\mathcal{P}$ is visible from a point in $S$. 
  The set $S$ is referred to as guards. 
  Assuming integer coordinates and a specific general position assumption, we  present the first $O(\log \text{OPT})$-approximation algorithm for the point guard problem.\footnote{For the benefit of the reviewers, we uploaded on youtube a video that describes informally the main ideas of the paper: \url{https://youtu.be/k5CDeimSuBM}.
  The video is meant as supplementary material and is not required for anything that is coming.}
  This algorithm combines ideas of a paper of Efrat and Har-Peled~\cite{DBLP:journals/ipl/EfratH06} and Deshpande et al.~\cite{DeshpandeThesis,DeshpandeWADS2007}. 
We also point out a mistake in the latter.
\end{abstract}

\section{Introduction}
  Given a simple polygon $\mathcal{P}$ on $n$ vertices, two points $x,y$ in $\mathcal{P}$ are said to be visible to each other if the line segment between $x$ and $y$ is contained in $\mathcal{P}$. The point-guard art gallery problem asks for a minimum set $S$ such that every point in $\P$ is visible from a point in $S$. 
  The set $S$ is referred to as guards.

  A huge amount of research is committed to the studies of combinatorial and algorithmic aspects of the art gallery problem, such as reflected by the following surveys~\cite{shermer1992recent, urrutia2000art, o1987art, ghosh2007visibility}. 
  Most of this research, however, is not focused directly on the art gallery problem but on variants, based on different definitions of visibility, restricted classes of polygons, different shapes  and positions of guards, and so on. 
  The most natural definition of visibility is arguably the one we gave above. 
Other possible definitions are: $x$ sees $y$ if the axis-parallel rectangle spanned by $x$ and $y$ is contained in $\mathcal{P}$; $x$ sees $y$ if the line segment $\seg(x,y)$ intersects $\mathcal{P}$ at most $c$ times, for some value of $c$; $x$ sees $y$ if there exists a straight-line path from $x$ to $y$ within $\mathcal{P}$ with at most $c$ bends.
  Common shapes of polygons include: simple polygons, polygons with holes, simple orthogonal polygons, $x$-monotone polygons and star-shaped polygons.  Common placements of guards include: vertex guards and point guards as defined above, but also edge-guard (guards are edges of the polygon), segment guards (guards are interior segments of the polygon) and perimeter guards (guards must be placed on the boundary of $\mathcal{P}$).  
The art gallery problem variants, also distinguish the way that the polygon is covered. For example, it might be required that every point is seen by two different guards; sometimes it might be required that every point is covered by one guard of each color; and recently the community gets also interested in conflict-free guard colorings. 
That is, every point has a color such that it sees exactly one guard of that color.
In 1978, Steve Fisk proved elegantly that $\lfloor n/3 \rfloor$ guards are always sufficient and sometimes necessary for a polygon with $n$ vertices~\cite{DBLP:journals/jct/Fisk78a}. 
Five years earlier, Victor Klee has posed this question to V\'aclav Chv\'atal, who soon gave a more complicated solution.
 This constitutes the first combinatorial result related to the art gallery problem. 
  
  On the algorithmic side, very few variants are known to be solvable in polynomial time~\cite{DBLP:journals/jcss/MotwaniRS90,durocher2013guarding} and most results are on approximating the minimum number of guards~\cite{DeshpandeThesis,DeshpandeWADS2007,ghosh2010approximation,DBLP:journals/comgeo/King13,DBLP:journals/dcg/Kirkpatrick15, DBLP:journals/ipl/EfratH06}.
  Many of the approximation algorithms are based on the fact that the range space defined by the visibility regions has bounded VC-dimension for simple polygons~\cite{valtr1998guarding,kalai1997guarding,gilbers2014new}. This makes it easy to use the algorithmic ideas of Clarkson~\cite{DBLP:conf/wads/Clarkson93,DBLP:journals/dcg/BronnimannG95}.

  On the lower bound side, Eidenbenz et al.~\cite{eidenbenz2001inapproximability} showed NP-hardness and inapproximability for most relevant variants. 
  In particular, they show for the main variants that there is no $c$-approximation algorithm for simple polygons, for some constant $c$.
  For polygons with holes, they can even show that there is no $o(\log n)$-approximation algorithm. 
  Also, their reduction from \textsc{Set-Cover} implies that the art gallery problem is W[2]-hard on polygons with holes and that there is no $n^{o(k)}$ algorithm, to determine if $k$ guards are sufficient, under the Exponential Time Hypothesis~\cite[Sec.4]{eidenbenz2001inapproximability}. 
Recently, a similar result was shown for simple polygons (without holes)~\cite{ESA-HARDNESS,DBLP:journals/corr/BonnetM16a}.
  
  Despite, the large amount of research on the art gallery problem, there is only one exact algorithmic result on
  the point guard variant.
  The result is not so well-known
  and attributed to Micha Sharir~\cite{DBLP:journals/ipl/EfratH06}:
  One can find in $n^{O(k)}$ time a set of $k$ guards for the point guard variant, if it exists.
  This result is quite easy to achieve with standard tools from real algebraic geometry~\cite{basu2011algorithms} and apparently quite hopeless to prove without this powerful machinery (see~\cite{belleville1991computing} for the very restricted case $k=2$). 
  Despite the fact that the algorithm uses remarkably sophisticated tools, it uses almost no problem-specific insights and no better exact algorithms are known. Some recent ETH-based lower bounds~\cite{ESA-HARDNESS,DBLP:journals/corr/BonnetM16a} suggest that there might be no better exact algorithm even for simple polygons.
  
  Regarding \emph{approximation} algorithms for the point guard variant, the results are similarly sparse. 
  For general polygons, Deshpande et al.\ gave a randomized pseudo-polynomial time $O(\log n)$-approximation algorithm~\cite{DeshpandeThesis, DeshpandeWADS2007}. However, we show that their algorithm is not correct.  Efrat and Har-Peled gave a randomized polynomial time $O(\log |OPT_{\text{grid}}|)$-approximation algorithm by restricting guards to a very fine grid~\cite{DBLP:journals/ipl/EfratH06}. However, they could not prove that their $S_{\text{grid}}$ grid solution is indeed an approximation of an optimal guard placement. 
Developing the ideas of Deshpande et al. in combination of the algorithm of Efrat and Har-Peled, we attain the first randomized polynomial-time approximation algorithm for simple polygons.
  Here, $OPT$ denotes an optimal set of guards and $OPT_{\text{grid}}$ an optimal set of guards that is restricted to some grid.
  At last, we want to mention that there exist approximation algorithms for monotone and rectilinear polygons~\cite{krohn2013approximate}, when the very restrictive structure of the polygon is exploited.
  
  To understand the lack of progress, note that the art gallery problem can be seen as a geometric hitting set problem.
  In a hitting set problem, we are given a universe $U$ and a set of subsets $S\subseteq 2^U$ and we are asked to find a smallest set $X \subseteq U$ such that $\forall r\in S \ \exists x\in X: x\in r$.
  Usually the set system is given explicitly or can be at least easily restricted to a set of polynomial size.
  In our case, the universe is the entire polygon (not just the boundary) and the set system is the set of visibility regions (given a point $x\in\P$, the visibility region $\vis(x)$ is defined as the set of points visible from $x$).
  The crucial point is that the set system is \emph{infinite}
  and no one has found a way to restrict the universe to a finite set (see~\cite{KnauerWitness, ayaz2015minimal} for some attempts).
  We also wish to quote a recent remark by Bhattiprolu and Har{-}Peled~\cite{DBLP:conf/compgeom/BhattiproluH16}:

  ''One of the more interesting versions of the geometric hitting set problem, is the art
gallery problem, where one is given a simple polygon in the plane, and one has to select a
set of points (inside or on the boundary of the polygon) that ``see'' the whole polygon. While
much research has gone into variants of this problem~\cite{o1987art}, nothing is known as far as an
approximation algorithm (for the general problem). The difficulty arises from the underlying
set system being infinite, see~\cite{DBLP:journals/ipl/EfratH06} for some efforts in better understanding this problem."

  Here, we present the first approximation algorithm for simple polygons under some mild assumptions.
  \begin{ass}[integer vertex representation]\label{ass:integer}
    Vertices are given by integers, represented in binary.
  \end{ass}
  An \emph{extension} of a polygon \P is a line that goes through
  two vertices of \P.
  \begin{ass}[general position assumption]\label{ass:generalPos}
    No three extensions meet in a point of $\P$ which is \emph{not} a vertex and no three vertices are collinear.
  \end{ass}
  Note that we allow that three (or more) extensions meet in a vertex or outside the polygon.
  \begin{theorem}\label{thm:ApproxAlgo}
Under Assumptions~\ref{ass:integer} and \ref{ass:generalPos}, there is a randomized $O(\log |OPT|)$-approximation algorithm for \pgag for simple polygons that runs in polynomial time in the size of the input.
  \end{theorem}

  The main technical idea is to show the following lemma:
  \begin{lemma}[Global Visibility Containment]\label{lem:GridReplacement}
  Let \P be some (not necessarily simple) polygon. 
  Under Assumptions~\ref{ass:integer} and \ref{ass:generalPos},
it holds that there exists a grid $\Gamma$ and a guard set $S_{\text{grid}}\subseteq \Gamma$, which sees the entire polygon and $|S_{\text{grid}}| = O(|S|)$, where $S$ is an optimal guard set.
\end{lemma}
  To be a bit more precise, let $M$ be the largest appearing integer. Then the number of points in $\Gamma$ is polynomial in $M$.
  This is potentially exponential in the size of the input.
  Thus algorithms that rely on storing all points of $\Gamma$ explicitly do not have polynomial worst case running time.
%
%
%
  The algorithm of Efrat and Har-Peled~\cite{DBLP:journals/ipl/EfratH06} does \emph{not} store every point of $\Gamma$ explicitly and, with the lemma above, the algorithm gives  an $O(\log |OPT|)$-approximation on the grid $\Gamma$.
  
%
  While Lemma~\ref{lem:GridReplacement} tells us that we can restrict our attention to a finite grid, when considering constant factor approximation, the same is not known for exact computation. 
In particular, it is not known whether the \pg problem lies in NP. 
  Recently, some researchers popularized an interesting class, called $\exists \R$, being somewhere between NP and PSPACE~\cite{Cardinal:2015:CGC:2852040.2852053, Schaefer2010, canny1988some, DBLP:journals/corr/Matousek14}.
  Many geometric problems, for which membership in NP is uncertain, have been shown to be complete for this class.
  This suggests that there might be indeed no polynomial sized witness for these problems as this would imply $NP = \exists \R$.
  The history of the art gallery problem suggests the possibility that the \pg problem is $\exists \R$-complete.
%
  If $NP\neq \exists \R$, then this would imply that there is indeed no hope to find a witness of polynomial size for the \pg problem.
  
In computational geometry and discrete geometry many papers assume that \emph{no three points lie on a line}. 
  Often this assumption is a pure technicality, in other cases the result might indeed be wrong without this assumption.
  In our case, we do believe that Lemma~\ref{lem:GridReplacement} could be proven without Assumption~\ref{ass:generalPos}, but it seems that some new ideas would be needed for a rigorous proof.
  See~\cite{DBLP:journals/siamdm/BaratDJPSSVW15} for an example where the main result is that some general position assumption can be weakened.
  The idea of general position assumptions is that a small random perturbation of the point set yields the assumption with probability almost $1$.
  In case that the points are given by integers small random perturbations, destroy the integer property. 
But random perturbations could be performed in a different way, by first multiplying all coordinates by some large constant $2^C \in \N$ and then add a random integer $x$ with $-C\leq x\leq C$.

  The integer representation assumption (Assumption~\ref{ass:integer}) seems to be very strong as it gives us useful distance bounds not just between any two different vertices of the polygon, but also between any two objects that do not share a point (see Lemma~\ref{lem:distances}).
  On the other hand, real computers work with binary numbers and cannot compute real numbers with arbitrary precision.
  The real-RAM model was introduced as a convenient theoretical framework to simplify the analysis of algorithms with numerical and/or geometrical flavors, see for instance~\cite[page $1$]{DBLP:conf/compgeom/KimR16} and \cite[Remark 23.1]{DBLP:journals/corr/KimR16}.
%
%
 Also note that Assumption~\ref{ass:integer} can be replaced by assuming that all coordinates are represented by rational numbers with specified nominator and denominator. (There could be other potentially more compact ways to specify rational numbers.)
 Multiplying all numbers with the smallest common multiple of the denominators takes polynomial time, makes all numbers integers and does not change the geometry of the problem.

  Given a polygon $\mathcal P$, we will always assume that all its vertices are given by \emph{positive} integers in binary. (This can be achieved in polynomial time.) We denote by $M$ the largest appearing integer and 
  we denote by $\diam(\P)$ the largest distance between 
  any two points in \P. Note that $\diam(\P)< 2 M$.
  We denote $L = 20 M > 10 $. Note that $\log L$ is linear in the input size.
  We define the grid 
  \[\Gamma = (L^{-\gridconstant}\cdot \mathbb{Z}^2) \cap \mathcal{P}.\] Note that all vertices of $\mathcal P$ have integer coordinates and thus are included in $\Gamma$.

\begin{theorem}[Efrat, Har-Peled~\cite{DBLP:journals/ipl/EfratH06}]
  Given a simple polygon $\mathcal P$ with $n$ vertices, one can spread a grid $\Gamma$ inside $\mathcal P$, and compute an $O(\log OPT_\textup{grid})$-approximation for the smallest subset of $\Gamma$ that sees $\mathcal P$. The expected running time of the algorithm is 
  \[O(n \, OPT_\textup{grid}^2  \log  OPT_{grid} \, \log( n\,  OPT_\textup{grid})   \, \log^2 \Delta),\]
  where $\Delta$ is the ratio between the diameter of the polygon and the grid size.
\end{theorem}
The term $OPT_{\textup{grid}}$ refers to the optimum, when restricted to the grid $\Gamma$. For the solution $S$ that is output by the algorithm of 
Efrat and Har-Peled holds $|S|= O(|OPT_{\textup{grid}}| \, \log |OPT_{\textup{grid}}|)$. However, Efrat and Har-Peled make no claim on the relation between $|S|$ and the actual optimum $|OPT|$.
Note that the grid size equals $w = L^{-\gridconstant}$, thus $\Delta \leq L^{\gridconstant + 1}=L^{12}$ and consequently $\log \Delta \leq 12 \log L$, which is polynomial in the size of the input. 

Efrat and Har-Peled implicitly use the real-RAM as model of computation: 
elementary computations are expected to take $O(1)$ time and coordinates of points are given by real numbers.
As we assume that coordinates are given by integers, the word-RAM or integer-RAM is a more appropriate model of computation. 
All we need to know about this model is that we can upper bound the time for elementary computations by a polynomial in the bit length of the involved numbers.
Thus, going from the real-RAM to the word-RAM only adds a polynomial factor in the running time of the algorithm of Efrat and Har-Peled.
Therefore, from the discussion above we see that it is sufficient to prove Lemma~\ref{lem:GridReplacement}.

\paragraph*{Organization.} 

In Section~\ref{sec:CE}, we will describe the counterexample to the algorithm of Deshpande {\em et. al.} This will be also very useful as a starting point of Section~\ref{sec:ProofIdea}, where we will give an elaborate overview of the forthcoming proofs.
Detailed and formal proofs are presented at Section~\ref{sec:formalProof}, where they can be read in logical 
order without references to Section~\ref{sec:ProofIdea}.
Finally in Section~\ref{sec:conclusion}, we briefly indicate the remaining open questions.
%

\section{Counterexample}\label{sec:CE}
    In this section, we will point out a mistake in the algorithm of Deshpande et al.~\cite{DeshpandeThesis,DeshpandeWADS2007}.
    This mistake though constitutes an interesting starting point for our purpose.
 
    The algorithm by Deshpande et al. can be described from a high level perspective as follows: maintain and refine a triangulation $T$ of the polygon until for every triangle $\Delta\in T$ holds the so called \emph{local visibility containment} property.
The local visibility containment property of $\Delta$ certifies that every point $p\in \Delta$ can only see points that are also seen by the vertices of $\Delta$.
    However, we will argue that it is impossible to attain the local visibility containment property with any finite triangulation; hence, the algorithm never stops.
    
Again, we want to mention that the paper of Deshpande et al. has ideas that helped to achieve the result of the present paper. 
In particular, we will show that the local visibility containment property does indeed hold most of the time.

    \begin{example}[Refutation of Deshpande et al.~\cite{DeshpandeThesis, DeshpandeWADS2007}]
    \begin{figure}[htbp]
      \centering
      \includegraphics{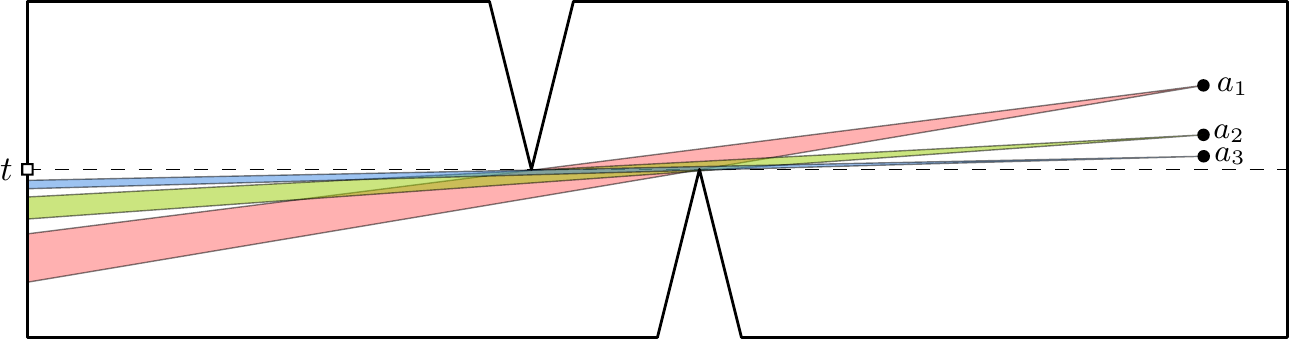}
      \caption{Illustration of the counterexample to the algorithm of Deshpande et al.~\cite{DeshpandeThesis, DeshpandeWADS2007}}
      \label{fig:DeshpandeCounterExample}
    \end{figure}
    See Figure~\ref{fig:DeshpandeCounterExample}, for the following description. We have two opposite reflex vertices with supporting line $\ell$. The points $(a_i)_{i\in \mathbb{N}}$ are chosen closer and closer to $\ell$ on the right side of the polygon. None of the $a_i$'s can see $t$, as this would require to be actually on $\ell$.
    Further, it is easy to see that the \emph{visible interval} next to $t$ gets smaller and smaller as well. 
In fact, we can choose the points $(a_i)_{i\in \mathbb{N}}$ in a way that their intervals will be all disjoint. 
    
    Consider now any finite collection of points $C$ in the vicinity of the $(a_i)_{i\in \mathbb{N}}$. We will show that there is some $a_i$, which sees some interval close to $t$, that is not seen by any point in $C$. Recall that no point sees the entire interval around $t$, but the visibility of the $a_i$'s come arbitrarily close to $t$. Thus, there is some $a_i$ that sees something that is not visible by any of the points of $C$. 
    \qed
\end{example}

\section{Proof Overview}\label{sec:ProofIdea}
In this section, we will describe all the proof ideas without going into too much technical details.
Definitions are given by figures and some technical conditions are not stated at all.
The reader who feels uncomfortable with this is deferred to Section~\ref{sec:formalProof}, where all definitions are rigorously made and all lemmas are presented with formal, detailed proofs.

Some readers might also be concerned with the wrong proportions of the figures. 
Some distances which are supposedly very small are displayed fairly large and vice versa.
The reader has to keep in mind that all figures only indicate principal behavior.
\bigskip 
 
  Our high level proof idea is that the local visibility property holds for every point $x$ that is far enough away from all extension lines.
  (Recall that the extension of two vertices is the line that contains these vertices.)
  In a second step, we will show that it is impossible to be close to more than $2$ extensions at the same time.
  We will add one vertex for each extension that $x$ is close two.
  Recall that vertices are also in $\Gamma$.

It turns out that the first step is considerably more tedious than the second one. The reason for that is that many of the elementary steps are not true in the naive way, one might think of at first. Thus we have to carefully define the sense in which they are true and handle the other case in a different manner.
The tricky bit is usually to identify these cases and to carefully define them. All proofs are elementary otherwise. 

\subsection{Benefit of integer coordinates}

The integer coordinate assumption implies not just that the distance between any two vertices is at least $1$ but it also gives useful lower bounds on distances between any two objects of interest that do not share a point. For example the distance between an extension $\ell$ and a vertex $v$ not on $\ell$ is at least $L^{-1}$.
Also the angle between any two non-parallel extensions is at least $L^{-2}$.
(Recall that $L$ is an upper bound on the diameter and the largest appearing integer.)
As these bounds are important for the intuition of the forthcoming ideas, we will proof one of them.
\begin{claim}
  Let $\ell = \ell(w_1,w_2)$ be the extension of the two vertices $w_1,w_2$ of \P and $v\notin \ell$ some other vertex of \P. Then, it holds $d =\dist(v,\ell) \geq L^{-1}$.
\end{claim}
Here $\dist(v,\ell)$ denotes the euclidean distance between $v$ and $\ell$.
\begin{proof}
\begin{figure}[htbp]
  \centering
  \includegraphics{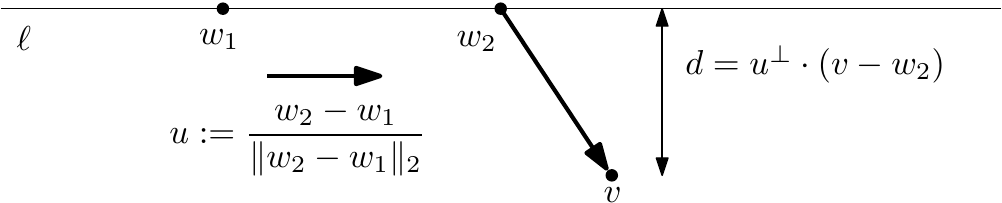}
  \caption{Computing the distance between a line and a vertex.}
  \label{fig:LineVertexDistance}
\end{figure}
  The distance $d$ can be computed as 
  \[d = \dfrac{|(v-w_1)\cdot (w_2-w_1)^{\bot}|}
  {\|w_2-w_1\|} \geq \frac{1}{\diam(\P)}\geq \frac{1}{L}.\]
  See Figure~\ref{fig:LineVertexDistance} for a way to derive this elementary formula.
  Here $\cdot$ denotes the scalar product and $x^\bot$ is the vector $x$ rotated by $90^\circ$ and $\Vert x\Vert_2$ is the euclidean norm of $x$.
  
  The key inside is that the nominator of this formula is at least $1$ as it is a non-zero integer by assumption. The denominator is upper bounded by the diameter of $\P$, which is in turn upper bounded by $L$.
\end{proof}
All other lower bounds are derived in the same spirit, however with worth bounds.
As we choose our grid width smaller than any of these bounds,
we will be in the very fortunate situation that everything looks very simple from a local perspective.

\subsection{Surrounding Grid Points}
  \begin{figure}[htbp]
  \centering
  \includegraphics{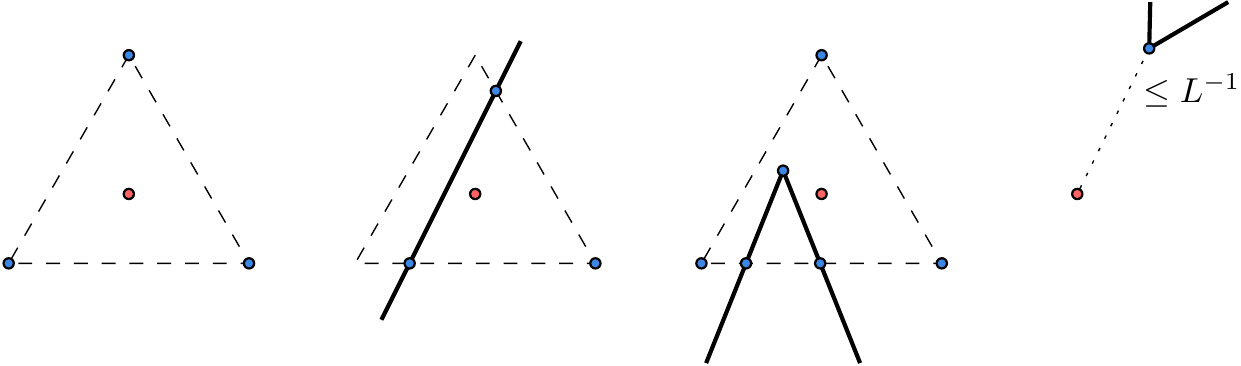}
  \caption{The red point indicates a point of the original optimal solution. 
  The blue points indicate the surrounding grid points that we choose.
  The polygon is indicated by bold lines.
  From left to right, we have the three cases: 
  interior case, boundary case and corner case.
  To the very right, we indicate that in every case vertices of \P with distance less than $L^{-1}$ are also included in $\grid^*(x)$.}
  \label{fig:GridTriangleFirst}
\end{figure}
  Given a point $x\in OPT$ we define $\grid(x)$ as some grid points around $x$, as displayed in Figure~\ref{fig:GridTriangleFirst}. The parameter $\alpha\ll 1$  determines how far away these points are. As the grid-width $w$ is much smaller than $\alpha$, we have a choice on how exactly to place these points.
  See Section~\ref{sec:AlphaGrid} for a precise definition.
  If there is any vertex $v$ of \P with $\dist(x,v)\leq  L^{-1}$, then this point will be included into $\grid^*(x)$.

\subsection{Local Visibility Containment}

\begin{figure}[htbp]
  \centering
    \begin{minipage}[b]{.55\linewidth}
      \centering      \includegraphics{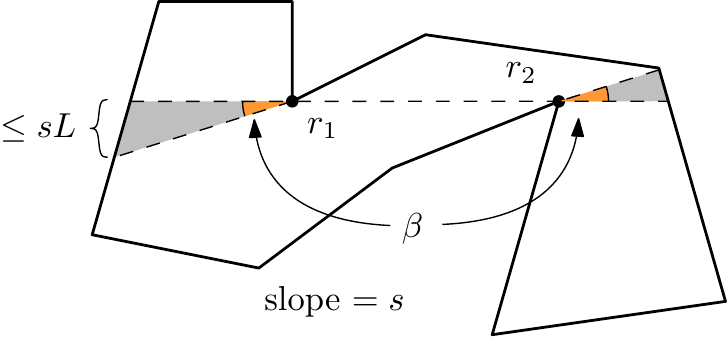}
      \subcaption{A polygon with two opposite reflex vertices and their $s$-bad region.}
      \label{fig:ThinBadRegionFirst}
    \end{minipage}%
    \hspace{0.2cm}
%
%
    \begin{minipage}[b]{.43\linewidth}
    \centering      \includegraphics[width = \textwidth]{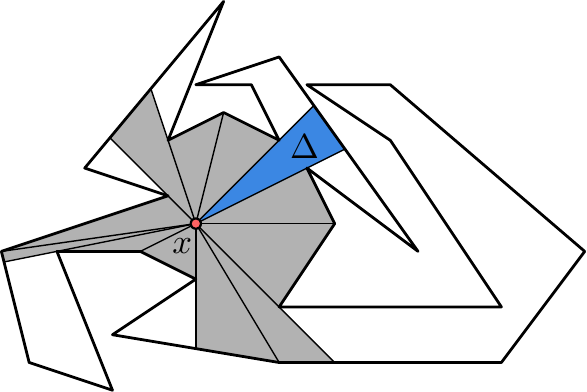}
    \subcaption{The star triangle decomposition of the visibility region of $x$.}
    \label{fig:TriangleDecompositionFirst}
    \end{minipage}%
  \caption{}
  \label{fig:ThinBadRegionsFirst}
\end{figure}

For any extension $\ell$ we define an $s$-bad region,
see the gray area in Figure~\ref{fig:ThinBadRegionFirst} for an illustration.
Note that the bad region consists of two connected components, each being a triangle.
The parameter $s = \tan(\beta)$ as indicated in the figure.
Further, for each point $x$, the visibility region can be decomposed into triangles as indicated in Figure~\ref{fig:TriangleDecompositionFirst}.

Let $\Delta$ be some triangle of the visibility region of $x$ (in blue in Figure~\ref{fig:TriangleDecompositionFirst}). Then the main lemma asserts that $\grid^*(x)$ sees $\Delta$ except, if $x$ is in an $s$-bad region of the vertices defining $\Delta$, see Lemma~\ref{lem:SpecialLocalVisibility}.
(With the right choice of $\alpha$ and $s$.)
Important is the one-to-one correspondence between the triangles that cannot be seen and the extension line that we can make responsible for it.

\begin{figure}[htb]
  \centering
    \includegraphics{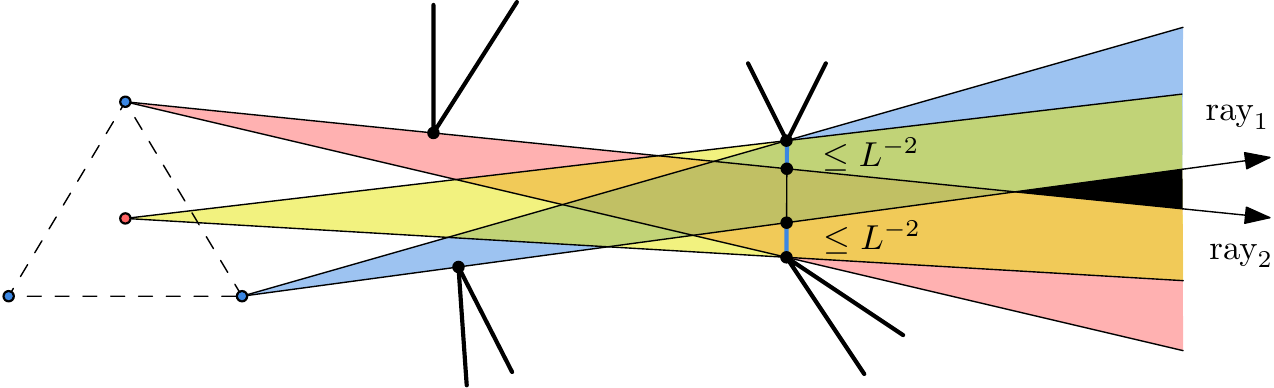}
  \caption{The visibility of the grid points $g\in \grid(x)$ can be blocked, but we can bound the amount by which it is blocked.
  The key idea to show that $R_2$ can be seen by $\grid(x)$ is to show that the region indicated in solid black is empty.}
  \label{fig:BlockingFirst}
\end{figure}

The first technical lemma concerns the possibility that
there is a reflex vertex $q$ that blocks the visibility of
the points of $\grid(x)$ onto $\Delta$, see Figure~\ref{fig:BlockingFirst}.
We can show that this can happen only in a negligible amount. 
For this recall that $\grid(x)$ is very close 
to $x$, but $q$ must have distance at least $L^{-1}$,
as otherwise it would be included into $\grid^*(x)$.
We call this phenomenon \emph{limited blocking}.

\begin{figure}[htb]
    \centering 
       \begin{minipage}[b]{.5\linewidth}
    \centering      \includegraphics{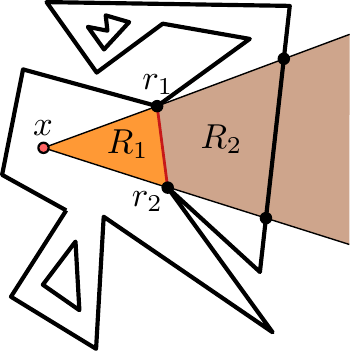}
    \subcaption{To show that each triangle of the visibility region is visible by $\grid^*(x)$, we treat the small triangle $R_1$ and the trapezoid $R_2$ individually. In particular, as we do not make use of the finiteness of $R_2$, we just assume it is an infinite cone.}
    \label{fig:SplitTriangle}
    \end{minipage}%
    \hspace{0.3cm}
     \begin{minipage}[b]{.45\linewidth}
    \centering      \includegraphics[width = \textwidth]{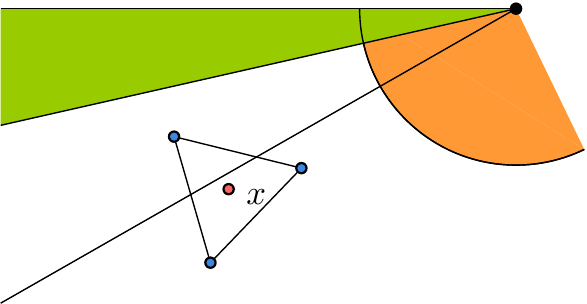}
    \subcaption{The point $x$ is not in an $s$-bad region implies that $\grid(x)$ is not in an $s/2$ bad region.}
    \label{fig:OutsideBadSimple}
    \end{minipage}%
    \caption{}
\end{figure}

Further the triangle $\Delta$ is split into a \emph{small triangle} ($R_1$)
and a trapezoid($R_2$), as indicated in Figure~\ref{fig:SplitTriangle}.
We show separately, for $R_1$ and $R_2$ that $\grid^*(x)$ sees these two regions.

To prove that $\grid^*(x)$ sees $R_1$ 
is not so difficult. We have already argued that
reflex vertices $v$ (with $\dist(x,v)\geq L^{-1}$) can only block a small part of the visibility of $\grid(x)$. 
(Recall that any vertex $v$ with $\dist(x,v)< L^{-1}$ is included in $\grid^*(x)$.)
With some simple case distinctions, we can argue that this visibility is sufficient to see $R_1$.
%
%
In particular, the argument does not rely on $x$ being outside a bad region.

To prove that $R_2$ can be seen by $\grid^*(x)$ is more demanding. 
As it seems not useful to use the boundedness of $R_2$, we just assume it to be an infinite cone and we try to show that $\grid^*(x)$ sees this cone.
Obviously, the part of $\partial\P$ ``behind'' $\seg(r_1,r_2)$ is not considered blocking.
The crucial step to show that $R_2$ can be seen by $\grid^*(x)$ is to show that the black region as indicated in Figure~\ref{fig:BlockingFirst} does not exist.
The idea is that this is implied if $\textup{ray}_1$ and $\textup{ray}_2$ diverge. 
In other words if $\textup{ray}_1$ and $\textup{ray}_2$ never meet and the black region is empty.
For this purpose, we make use of the fact that $\dist(g_1,g_2)\approx \alpha$, for any $g_1,g_2\in \grid(x)$ by definition, while $\dist(r_1,r_2) \geq 1$, because of integer coordinates. 
Thus intuitively, the distance of $\ray_1$ and $\ray_2$ is closer at its apex than at the segment $\seg(r_1,r_2)$.
For the proof to gp through we further need $\grid^*(x)$ not to be in an $s/2$-bad region, which follows from $x$ not being in an $s$-bad region, see Figure~\ref{fig:OutsideBadSimple}.
The proof becomes technically even more tedious as we have to take into consideration that the visibility of $\grid^*(x)$ might be partially blocked.
This also forces us to introduce embiggened bad regions, which differ only marginally from bad regions as defined above.

\subsection{Global Visibility Containment}
\begin{figure}[htbp]
  \centering
    \includegraphics{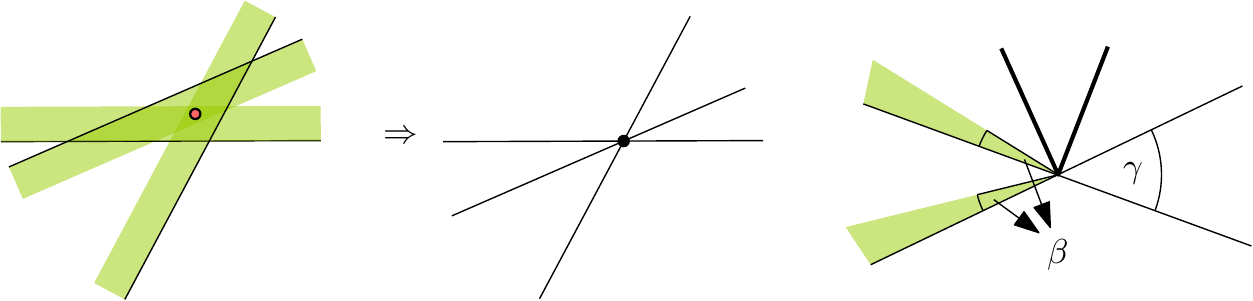}
  \caption{Three bad regions meeting in an interior point implies that the extensions must meet in a single point. No two bad regions intersect in the vicinity of a vertex, as they are defined by some angle $\beta \ll L^{-2}$.
  But the angle $\gamma$ between any two extensions is at least $L^{-2}$.}
  \label{fig:NoThreeBadFirst}
\end{figure}
Given a minimum solution $OPT$, we describe a set $G\subseteq \Gamma$ of size $O(|OPT|)$. We will also show that $G$ sees the entire polygon.
For each $x\in OPT$, $G$ contains $\grid^*(x)$. Further if $x$ is contained in an $s$-bad region, $G$ contains at least one of the vertices defining it.
It is clear by the previous discussion that $G$ sees the entire polygon, 
as the only part that is not seen by $\grid^*(x)$ are some small regions, which are entirely seen by the vertices bounding it.

It remains to show that there is no point in three bad regions.
For this, we heavily rely on the integer coordinates and the general position assumption. 
Note that the integer coordinate assumption implies not just that the distance between any two vertices is at least $1$ but also that the distance between any extension $\ell$ and a vertex $v$ not on $\ell$ is at least $L^{-1}$.
Also the angle between any two extensions is at least $L^{-2}$.
(Recall that $L$ is an upper bound on the diameter and the largest appearing integer.)
These bounds and other bounds of this kind imply that 
if any three bad regions meet in the interior, then their extension lines must meet in a single point, see Figure~\ref{fig:NoThreeBadFirst}. We exclude this by our general position assumption. Close to a vertex, we use a different argument:
No two bad regions intersect in the vicinity of a vertex, as bad regions are defined by some angle $\beta$ with $\tan(\beta) \ll L^{-2}$.
But the angle $\gamma$ between any two extensions is at least $L^{-2}$.

This implies that each $x$ is in at most $2$ bad regions and $|G| = (7+2)|OPT| = O(|OPT|)$.

  \begin{figure}[htbp]
  \centering
    \includegraphics{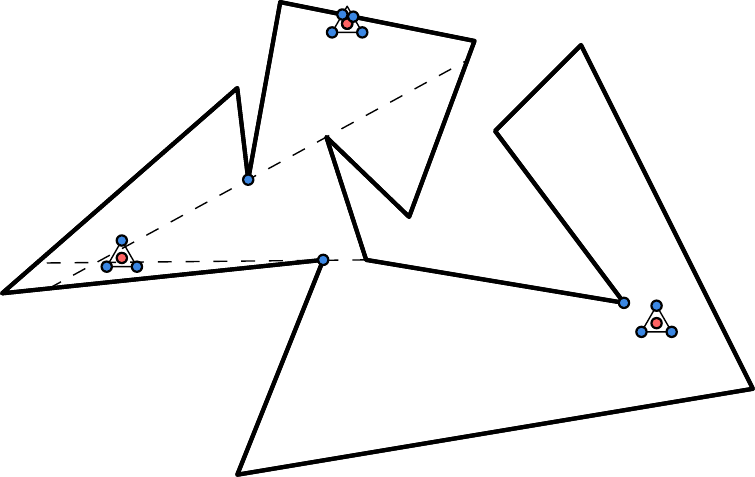}
  \caption{The red dots indicate the optimal solution. The blue dots indicate the  
  The red dot on the top is in the interior case and four grid points are added around it.
  The red dot on the left is too close to two supporting lines and we add one of the reflex vertices of each of the supporting lines. 
  The red dot to the right has distance less than $L^{-1}$ to a reflex vertex, so we add that vertex as well to $G$.}
  \label{fig:GlobalVisbilityFirst}
\end{figure}

\section{Proofs} \label{sec:formalProof} 
\subsection{Preliminaries} \label{sec:prelim}


\emph{Polygons and visibility.}
For any two distinct points $v$ and $w$ in the plane, 
we denote by $\seg(v,w)$ the segment whose two endpoints 
are $v$ and $w$, by $\ray(v,w)$ the ray starting at $v$ 
and passing through $w$, by $\ell(v,w)$ the supporting 
line passing through $v$ and $w$. 
We think of $\ell(v,w)$ as directed from $v$ to $w$.
Given a \emph{directed} line $\ell$, 
we denote by $\ell^+$ the half-plane to the right 
of $\ell$ bounded by $\ell$.
We denote by $\ell^-$ the half-plane to the left 
of $\ell$ bounded by $\ell$.
We also denote by $\disk(v,r)$ the disk 
centered in point $v$ and whose radius is $r$, 
and by $\dist(a,b)$ the distance between 
point $a$ and point $b$.
We denote by $\diam(\P)$ the diameter of \P and by $M$
the largest appearing integer of any vertex of $\P$.
We further assume that all coordinates are represented by
positive integers. 
(As already stated earlier, this can be achieved in polynomial time.)
We define $L= 20M$.
This implies $\diam(\P) \leq 2M$ and $10\, \diam(\P) \leq L$.
A polygon is \emph{simple} if it is not self-crossing and has no holes.
For any point $x$ in a polygon $\P$, $\vis(x)$, denotes the \emph{visibility region} of $x$ within $\mathcal P$, that is the set of all the points $y \in \mathcal P$ such that segment $\seg(x,y)$ is entirely contained in $\mathcal P$.

\begin{figure}[hbt]
\centering
\begin{tabular}{ |c | c|  }
  \hline
  symbol & definition \\
  \hline
  \P & denotes underlying polygon  (we assume all coordinates to be positive) \\
  $\vis(x)$ & visibility region of $x$\\
  $\diam(\P)$ & the largest distance between any two points in \P \\
  $M$ & largest appearing integer of \P \\
  L & = $20M$ (it holds $\diam(\P)\leq 10L$.) \\
  $\Gamma$ & the grid $(L^{-11}\cdot \mathbb{Z}^2) \cap \mathcal{P}$.\\
  $\seg(v,w)$ & segment with endpoints $v$ and $w$  \\
  $v(p,q)\in S^1$  & direction from $p$ to $q$ \\
  $\ray(p,q)$  & $p,q$ points in the plane: ray with apex $p$ in direction $v(p,q)$ \\
  $\ray(p,v)$  & $p$ points in the plane, $v\in S^1$ direction: ray with apex $p$ in direction $v$ \\
  $\ell(v,w)$ & line through $v$ and $w$ directed from $v$ to $w$\\
  $\ell^+, \ell^-$ & the half-plane to the right respectively to the left of line $\ell$\\
  $\text{disk}(v,r)$ &  disk 
centered in point $v$ with radius  $r$\\
$\text{dist}(a,b)$ & euclidean distance between $a$ and $b$.\\
$\text{dist}_{\textrm{horizontal}}(a,b)$ & $|a_x- b_x|$\\
$\text{dist}_{\textrm{vertical}}(a,b)$ & $|a_y- b_y|$\\  
%
   \grid(x) & some grid points around $x$, see Figure~\ref{fig:GridTriangle}\\
   \grid$^*$(x) &  $\grid(x)$ and including a possible vertex $v$ with $\dist(r,x)\leq L^{-1}$\\
     $\cone(x,r_1,r_2)$ & a cone with apex $x$ bounded by $\ray(x,r_1)$ and $\ray(x,r_2)$ \\
  $\cone(x)$ & we use this notation, when $r_1$ and $r_2$ are clear from context. \\
  $\apex(x,y)$ & see Figure~\ref{fig:ConeCutting} and Definition~\ref{def:ConeProperty} \\
  \hline
\end{tabular}
\caption{This is a glossary of all used notation. Not all of which is introduced in the Preliminaries.}
\end{figure}

\subsection{Benefit of Integer Coordinates} \label{sec:integer}

The way we use the integer coordinate assumption is to infer distance lower bounds between various objects of interest. 
The proofs are technical and not very enlightening. 
  Fortunately they are short.

\begin{restatable}[Distances]{lem}{LemDistances}\label{lem:distances}
  Let $\mathcal P$ be a polygon with integer coordinates and $L$ as defined above.
  Let $v$ and $w$ be vertices of $\mathcal P$, $\ell$ and $\ell'$ supporting lines of two vertices, and $p$ and $q$ intersections of supporting lines.
  \begin{enumerate}
   \item \label{itm:VertexVertex} 
   $\dist(v,w)>0 \Rightarrow \dist(v,w)\geq 1$.
   \item \label{itm:VertexLine}
   $\dist(v,\ell) >0 \Rightarrow \dist(v,\ell) \geq L{-1}$.
   \item \label{itm:IntersectionLine}
   $\dist(p,\ell) >0 \Rightarrow \dist(p,\ell) \geq L^{-5}$.
   \item \label{itm:IntersectionIntersection}
   $\dist(p,q) >0 \Rightarrow \dist(p,q) \geq L^{-4}$.
   \item \label{itm:parralelLines} Let $\ell\neq \ell'$ be parallel. Then $\dist(\ell,\ell') \geq L^{-1}$.
   \item \label{itm:slope} Let $\ell\neq \ell'$ be any two non-parallel supporting lines and $\alpha$ the smaller angle between them. Then holds $\tan(\alpha)\geq 8L^{-2}$.
   \item \label{itm:LinesCommonIntersection} Let $a\in \P$ be a point and $\ell_1$ and $\ell_2$ be some non-parallel lines with $\dist(\ell_i, a )< d$, for $i = 1,2$. Then $\ell_1$ and $\ell_2$ intersect in a point $p$ with $\dist(a,p)\leq dL^2$.
  \end{enumerate}
\end{restatable}

  \begin{proof}
  Case~\ref{itm:VertexVertex} seems trivial, but follows the same general principal as the other bounds.
  All these distances, are realized by geometric objects, and these geometric objects are represented with the help of the input integers.
  In order to compute theses distances some elementary calculations are performed and solutions can be expressed as fractions of the input integers. Using the lower bound one and the upper bound $L$ or $L/2$ on these integers or derived expressions give the desired results.
  
  Consider Case~\ref{itm:VertexLine}. Let $\ell$ be the supporting line of $w_1$ and $w_2$. 
  The distance $d$ between $v$ and $\ell$ can be expressed as 
  $\tfrac{|(v-w_1)\cdot (w_2-w_1)^{\bot}|}{\|w_2-w_1\|}$, where $\binom{x_1}{x_2}^{\bot} = \binom{x_2}{-x_1}$ represents the orthogonal vector to $x$ and $\cdot$ indicates the dot product. 
  As $v$ is not on $\ell$ the nominator is lower bounded by one. The denominator is upper bounded by the 
  diameter $\diam(\P)$.
  
  Consider Case~\ref{itm:IntersectionLine}. Let $p$ be the intersection of $\ell_1 = \ell(v_1,v_2)$ and $\ell_2 = \ell(u_1,u_2)$. Then $p$ is the unique solution to the linear equations 
  $(u_2-u_1) \cdot (p-u_1)=0 $ and 
  $(v_2-v_1) \cdot (p-v_1)=0 $.
  So let us write this as an abstract linear equation $A\cdot x = b$.
  By Cramer's rule holds $x_i = \tfrac{\det(A_i)}{\det(A)}$, for $i = 1,2$.
  Here $\det(.)$ is the determinant and $A_i$ is the matrix with the $i$-th column replaced by $b$. It is easy to see that $det(A)$ is bounded by $L^2$. 
  Let $\ell=\ell(w_1,w_2)$ be a different line.
  Observe that the points $w_1$, $w_2$ and $p$ lie on a grid $\Gamma^*$ of width $1/\det(A)$. 
  After scaling everything with $det(A)$ the diameter 
  $\diam(\P)$ 
of this grid becomes $det(A) L \leq L^3$. Thus by Case~\ref{itm:VertexLine} the distance between the line $\ell$ and the grid point $p$ is lower bounded by $(1/L^*) \geq 1/L^3$. Scaling everything back by $1/det(A)\geq L^{-2}$ yields the claimed bound.
  
  Consider case~\ref{itm:IntersectionIntersection}. 
  Let $p$ be the intersection of $\ell_1 = \ell(v_1,v_2)$ and $\ell_2 = \ell(u_1,u_2)$.
  And likewise let $q$ be the intersection of $\ell_3 = \ell(w_1,w_2)$ and $\ell_2 = \ell(t_1,t_2)$. Then $p$ is the unique solution to the linear equations 
  $(u_2-u_1) \cdot (p-u_1)=0 $ and 
  $(v_2-v_1) \cdot (p-v_1)=0 $.
  So let us write this as an abstract linear equation $A\cdot x = b$.
  By Cramer's rule holds $x_i = \tfrac{det(A_i)}{det(A)}$, for $i = 1,2$.
  Thus the coefficients of $p-q$ can be represented by $\tfrac{a}{det(A)\cdot det(A')}$ for some value $a\neq 0$ and some matrices $A$ and $A'$. Note that $det(A)$ and $det(A')$ are bounded by   $L^2$ from above.

  Case~\ref{itm:parralelLines} follows from Case~\ref{itm:VertexLine}. Let $\ell'=\ell(w_1,w_2)$. Then $\dist(\ell,\ell') = \dist(\ell,w_1)\geq 1/L$.
  
  Consider case~\ref{itm:slope}. We assume $\ell = \ell(v_1,v_2)$ and $\ell'=\ell(w_1,w_2)$. The expressions $\tan(\alpha)$ can be computed by \[\frac{\|(v_2-v_1)\times (w_2-w_1)\|}{\| (v_2-v_1) \|\cdot \| (w_2-w_1) \|} \geq \frac{1}{\diam(\P)^2} \geq \frac{8}{L^2}.\] 
  

  \begin{figure}[htbp]
    \centering
    \includegraphics{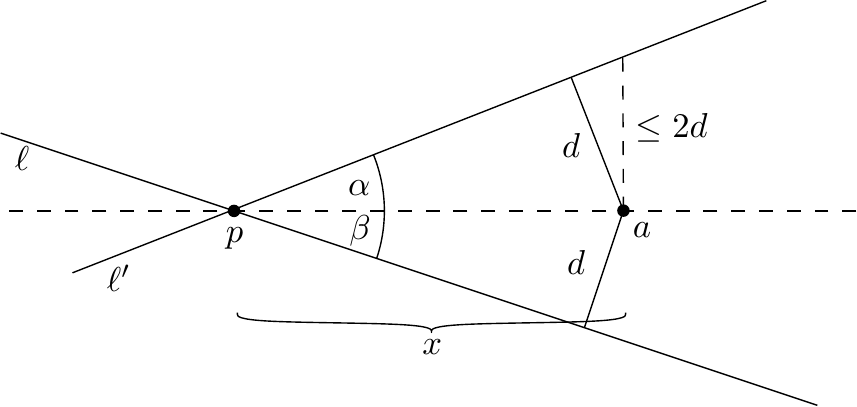}
    \caption{}
    \label{fig:NonParralelLines}
  \end{figure}

  For case~\ref{itm:LinesCommonIntersection}, see Figure~\ref{fig:NonParralelLines} and the notation therein. 
  Without loss of generality we assume $\alpha\geq \beta$. 
  It follows from elementary calculus that $z \leq \tan(z) \leq 2z$, for any sufficiently small $z$.  
  This implies
  $8L^{-2} \leq \tan (\alpha + \beta) \leq 2(\alpha + \beta) \leq 4\alpha \leq 4 \tan \alpha$.
  We follow from \,$x \tan\alpha \leq 2 d $, it holds $x \leq 2d / \tan \alpha \leq d L^2$.
\end{proof}

\subsection{Defining surrounding grid points}\label{sec:AlphaGrid}

\begin{figure}[htbp]
  \centering
  \includegraphics{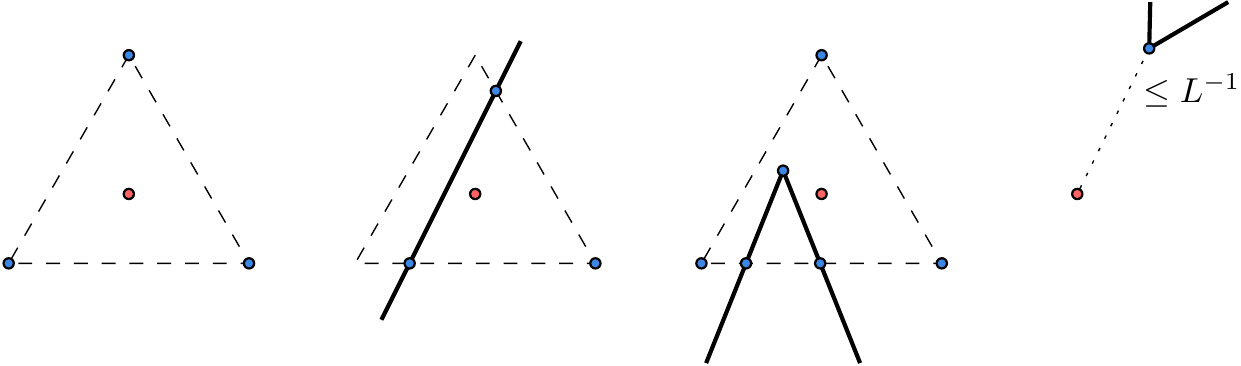}
  \caption{The red point indicates a point of the original optimal solution. 
  The blue points indicate the surrounding grid points that we choose.
  The polygon is indicated by bold lines.
  From left to right, we have the three cases: 
  interior case, boundary case and corner case.
  To the very right, we indicate that in every case vertices of \P with distance less than $L^{-1}$ are also included in $\grid^*(x)$.}
  \label{fig:GridTriangle}
\end{figure}

The point of this section is to define for a point $x$ 
some grid-points $\grid(x)$ surrounding $x$. Morally, $\grid(x)$ should see
whatever $x$ sees. However, this cannot be achieved in general.

\begin{definition}[Rounding] \label{def:Rounding}
Given a point $x \in \P$, we define the point 
$\round(x) = g\in \Gamma\subset \P$ as the 
closest grid-point to $x$. 
In case that there are several grid-points 
with the same minimum distance to $x$, 
we choose the one with lexicographic 
smallest coordinates. 
The important point here is that $\round(x)$ is uniquely defined.
\end{definition}

\begin{definition}[Surrounding Grid points] \label{def:SurroundingGrid}
Given a point $x\in \P$ and a number $\alpha \in [0,L^{-2}]$, we define  
$ \grid(x)$ as a set of grid points around $x$. 
The definition depends on the position of $x$ and the value $\alpha$.
Let $c$ be a circle with radius alpha and center $x$. Then there exists a unique
triangle $\Delta(x)$ inscribed $c$ such that the lower side of 
$\Delta(x)$ is horizontal.
We distinguish three cases. 
In the \emph{interior} case, $\Delta(x)$ and $\partial \P$ are disjoint. 
In the \emph{boundary} case, $\Delta(x)$ and $\partial \P$ have a non-empty intersection, but no vertex of \P is contained in $\Delta(x)$.
In the \emph{corner} case, on vertex of \P is contained in $\Delta$.
It is easy to see that this covers all the cases.
We also say a point $x$ in the interior case, and so on.
In the \emph{interior case} $ \grid(x)$ is defined as follows.
Let $v_1,v_2,v_3$ be the vertices of $\Delta$. Then the grid points $\round(v_i)$, for all $i = 1,2,3$ form the surrounding grid points.
In the \emph{boundary case} $ \grid(x)$ is defined as follows.
Let $S$ be the set of vertices of $\Delta$ and all intersection points of $\partial \P$ with $\partial \Delta(x)$. Then the grid points $\round(v)$, for all $v\in S\cap \P$ form the surrounding grid points.
In the \emph{corner case} $ \grid(x)$ is defined as follows.
Let $S$ be the set of vertices of $\Delta$. all intersection points of $\partial \P$ with $\partial \Delta$ and the vertex of \P contained in $\Delta$. Then the grid points $\round(v)$, for all $v\in S\cap \P$ form the surrounding grid points.
In any case, if there is a reflex vertex $r$ with $\dist(x,r)\leq L^{-1}$ then we include $r$ in the set $ \grid^*(x)= r \cup  \grid(x)$ as well.
We will usually denote the points in  $\grid(x)$ with $g_1,g_2$ or just $g$.
\end{definition}

As we will choose $\alpha \gg w$, the difference between $x$ and $\round(x)$ is negligible and thus we will assume that $x = \round(x)$.

\subsection{Local Visibility Containment}

\begin{definition}[Local Visibility Containment property]
 We say a point $x$ has the 
 \emph{$\alpha$-local visibility containment property} 
 if \[\vis(x)\subseteq \bigcup_{g\in \grid^*(x)}\, \vis(g).\] 
\end{definition}

\begin{figure}[htbp]

  \centering
    
    \begin{minipage}[b]{.5\linewidth}
    \centering      \includegraphics{ThinBadRegions}
    \subcaption{A polygon with two opposite reflex vertices and their $s$-bad regions.}
    \label{fig:NormalBadRegion37}
    \end{minipage}%
    \begin{minipage}[b]{.5\linewidth}
    \centering      \includegraphics[width = .9\textwidth]{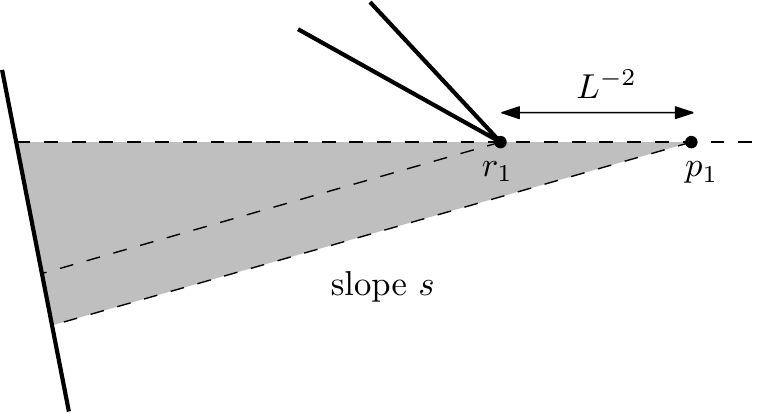}
    \subcaption{The embiggened $s$-bad region.}
    \label{fig:EmbiggenedBadRegion}
    \end{minipage}%

  \caption{Illustration of bad regions and embiggened bad regions.}
  \label{fig:ThinBadRegions}
\end{figure}

\begin{definition}[Opposite reflex vertices and bad regions] \label{def:BadRegion}
  Given a polygon $\mathcal P$ and two reflex vertices $r_1$ and $r_2$, consider the supporting line $\ell = \ell(r_1,r_2)$ restricted to $\mathcal P$. 
  We say $r_1$ is \emph{opposite} to $r_2$ if 
  both incident edges of $r_1$ lie on the opposite side of 
  $\ell$ as the edges of $r_2$,
%
  see Figure~\ref{fig:NormalBadRegion37}, for an illustration.
  Given two opposite reflex vertices $r_1$ and $r_2$, we define their \emph{$s$-bad region} as the union of the two triangles as in Figure~\ref{fig:NormalBadRegion37}, where we set the slope to $s$.
  Alternatively, let $\beta$ be the angle at $r_i$, then we define 
  $\tan(\beta) = s$.
  
  We denote by $p_i$ the point on $\seg(r_1,r_2)$ with $\dist(p_i,r_i) = L^{-2}$, for $i=1,2$.
  We define the \emph{embiggened $s$-bad region} of $r_1$ and $r_2$ as the union of the two triangles $\Delta_1$ and $\Delta_2$, one
  of which is displayed  in Figure~\ref{fig:EmbiggenedBadRegion}.
  Formally triangle $\Delta_i$ has one corner $p_i$. The two edges
  incident to $p_i$ form a slope of $s$. One of these edges is part of $\ell(r_1,r_2)$ the other is in the interior of \P. 
  The last remaining edge is part of $\partial \P$.
\end{definition}
  
\begin{figure}[htbp]
    \centering 
    \begin{minipage}[b]{.45\linewidth}
    \centering      \includegraphics{TriangleDecomposition}
    \subcaption{The star triangle decomposition of the visibility region of $x$.}
    \label{fig:TriangleDecomposition}
    \end{minipage}%
  \hspace{0.2cm}
    \begin{minipage}[b]{.45\linewidth}
    \centering      \includegraphics{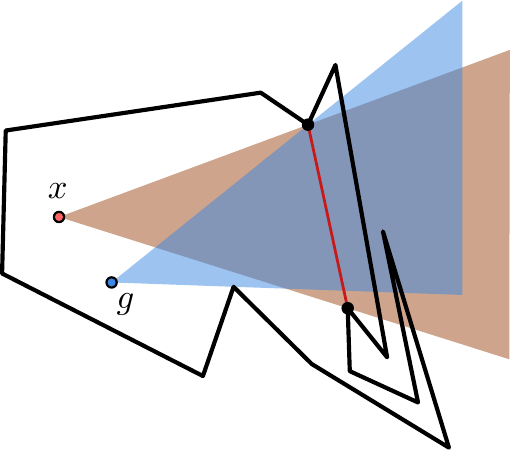}
    \subcaption{The cone of $x$ and $g$ are displayed. Both cones are unbounded. The cone of $g$ is not bounded by the vertices that define it.}
    \label{fig:GridCone}
    \end{minipage}%
    \caption{Illustration of the star triangle decomposition and cones.}
	\label{fig:VisibilityRegionDecomposition}
\end{figure}

\begin{definition}[Triangle decomposition of visibility regions and cones]\label{def:VisibilityCones}
  Given a polygon \P and some point $x$, we define the \emph{star triangle decomposition} of $\vis(x)$ as follows. Let $v_1,\ldots,v_t$ be the vertices visible from $x$ in clockwise order. Then any two consecutive vertices $u,v$ together with $x$ define a triangle $\Delta$ of $\vis(x)$. Note that $u,v$ are not necessarily the vertices of $\Delta$, see Figure~\ref{fig:TriangleDecomposition}. 
  Also note that the definition does not require the polygon to be simple.
  (The polygon could have holes.)
  
  We denote by $\cone(x,u,v) = \cone(x)$ the cone with apex $x$ that is bounded by $\ray(x,u)$ and $\ray(x,v)$, see Figure~\ref{fig:GridCone}. We will assume that $u$ and $v$ are implicitly known and omit to mention them if there is no ambiguity. Note that $\cone(x)$ is unbounded and not contained in \P.

  
   Let $u,v$ be vertices, not necessarily 
  visible from some other point $g$.
  Let $s\subseteq \seg(u,v)$, the part of the segment that is 
  visible from $g$. Then we define 
  \[\cone(g) = \bigcup_{t\in s} \ray(g,t).\]
  See the blue cone in Figure~\ref{fig:GridCone}.
  We say $\grid(x)$ \emph{sees} $\cone(x)$ if 
  \[\cone(x) \subseteq \bigcup_{g\in \grid(x)} \vis(g) \quad  \cup \quad
  \bigcup_{g\in \grid(x)}\cone(g).\]
  Note that $\cone(g)$ is not contained into $\vis(g)$ as the cone is unbounded, whereas the visibility region is contained in \P.
  We define $\grid^*(x)$ \emph{sees} $\cone(x)$ in the same fashion.  
\end{definition}

The purpose of this Section is to prove Lemma~\ref{lem:SpecialLocalVisibility}.

\begin{restatable}
[Special Local Visibility Containment Property]
{lem}{SpecialVisa}
\label{lem:SpecialLocalVisibility}
  Let  $r_1$ and $r_2$ be two consecutive vertices
  in the clockwise order of the vertices visible from $x \in \P$
  and let $x$ be outside the $s$-bad region of 
  the vertices $r_1$ and $r_2$.
  We make the following assumptions:
  \begin{enumerate}[noitemsep,topsep=0pt,parsep=0pt,partopsep=0pt]
   \item $s\leq L^{-3} $
   \item $\alpha \leq L^{-7}$.
   \item $16L\alpha \leq s$
  \end{enumerate}
%
  Then $\grid^*(x)$ sees $\cone(x)$.
\end{restatable}
Note that, we do not exclude that $x$ is outside \emph{any} $s$-bad region
but only outside the bad region of $r_1$ and $r_2$. This subtlety makes the proof more complicated, as it might be that $x$ is in an $s$-bad region with respect to some other pair of reflex vertices. For instance $q,r_1$, could be such a pair. It is not difficult to construct an example such that $q$   blocks part of the visibility of $\grid(x)$.
The proof idea is to show that this blocking is very limited (see Lemma~\ref{lem:LimitedBlocking}) and then use only this partially 
obstructed visibility.

Seeing $\cone(x)$ implies obviously also that the triangle
$\Delta\subseteq \cone(x)$ is seen by $\grid(x)$.
An immediate consequence is the following nice lemma.

\begin{lem}[Local Visibility Containment Property]\label{lem:LocalVisibility}
  Every point $x\in \P$ outside any $s$-bad region
  has the $\alpha$-local visibility containment property.
  We make the following assumptions:
  \begin{enumerate}[noitemsep,topsep=0pt,parsep=0pt,partopsep=0pt]
   \item $s \leq L^{-3} $
   \item $\alpha \leq L^{-7}$.
   \item $16\, L\,\alpha \leq s$
  \end{enumerate}
\end{lem}

\begin{proof}
  By Lemma~\ref{lem:SpecialLocalVisibility}, every triangle $\Delta$ defined by the star triangle decomposition of $\vis(x)$ is seen by $\grid^*(x)$.
\end{proof}

\begin{figure}[htbp]
    \centering 
     \begin{minipage}[b]{\linewidth}
      \centering 
      \includegraphics{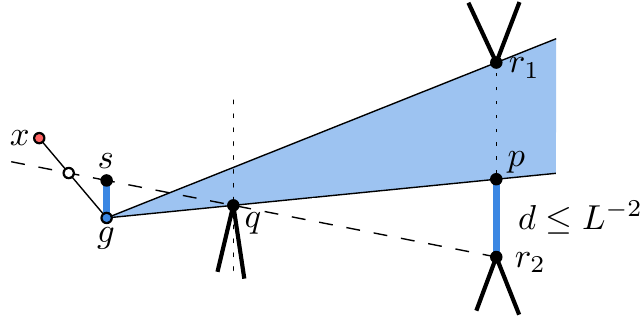}
      \subcaption{The reflex vertex $q$ blocks the visibility of $g$.
      Fortunately, we can show that only a very small part of the visibility is actually blocked.}
	\label{fig:LimitedBlocking}
     \end{minipage} 
	
	\begin{minipage}[b]{\linewidth}
      \centering 
      \includegraphics{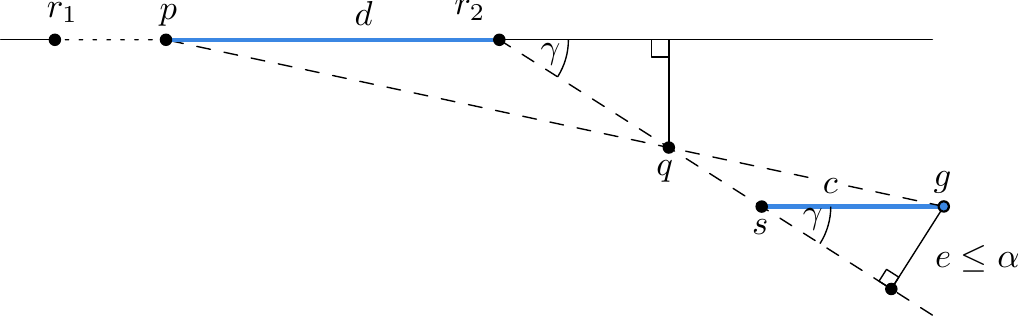}
	  \subcaption{This Figure illustrates the proof of Lemma~\ref{lem:LimitedBlocking}.}
      \label{fig:BlockedCaseA}
     \end{minipage} 
      \caption{}
 	\label{fig:unnamed}
\end{figure}

For Lemma~\ref{lem:LimitedBlocking}, see Figure~\ref{fig:LimitedBlocking} for an illustration of the setup.
The next Lemma deals with the very special situation that
there exists a reflex vertex $q$ blocking the visibility of 
some $g\in \grid(x)$, although $q\notin \cone(x)$.
The critical assumption is $\dist(q,x)\geq L^{-1}$.

\begin{restatable}[Limited Blocking]{lem}{LimitedBlocking}\label{lem:LimitedBlocking}
  Let $x\in \P$, let $r_1,r_2$  be two vertices and $g \in \grid(x)$ such that $g\notin \cone(x)$ and $\alpha\leq L^{-7}$.
  Further, let $q \in \cone(g)$ be a reflex vertex with $\dist(x,q)> L^{-1}$.
  We denote by $p$ the intersection point 
  between $\ell(g,q)$ and $\seg(r_1,r_2)$. 
  Let $g$ and $r_2$ be on the same side of $\cone(x)$.
  Then holds: \[d = \dist(p,r_2) \leq L^{-2} 
  .\]
\end{restatable}

\begin{proof}
  We show first $\dist(g, \ell(r_1,r_2)) \geq L^{-1}$. 
  Let $\ell$ be the line parallel to $\ell(r_1,r_2)$ containing the point $q$.
  Clearly, these two lines have distance at least $L^{-1}$, by Lemma~\ref{lem:distances} Item~\ref{itm:VertexLine}. And thus also $g$
  has at least this distance to $\ell(r_1,r_2)$, see Figure~\ref{fig:LimitedBlocking}.

  For the other part of the proof let us assume that $\ell(r_1,r_2)$ is vertical.
  We denote by $c$ the \emph{vertical} distance between $g$ and $\ell(q,r_2)$
  and by $s$ the point realizing this distance, see Figure~\ref{fig:LimitedBlocking}. 
  Note that $\dist(g,q) \geq \dist(x,q) - \dist(x,g) \geq L^{-1} - \alpha \geq 1/(2 L).$
  By Thales Theorem we know:
  \[\frac{d}{\dist(p,q)} 
  = \frac{c}{\dist(g,q)}
  \leq \frac{c}{1/(2L)}
  = 2c L.\] This implies \[d\leq 2c L \, \dist(p,q)\leq  2c L^{2} \]
  
  It is sufficient to show $c\leq  \alpha L^2$, as $\alpha \leq L^{-7}$.
  For this purpose, we rotate $\ell(r_1,r_2)$ again so that it becomes horizontal, see Figure~\ref{fig:BlockedCaseA}.
  Now $c$ is the vertical distance between $g$ and $\ell(q,r_2)$.
  We define $e = \dist(g,\ell(r_2,q))$ and
  we denote by $\gamma$ the angle at $r_2$ formed by $\ell(r_2,q)$ and $\ell(r_1,r_2)$. We consider here only the case that $\gamma < 90^{\circ}$, the other case is symmetric.
  We note that $\gamma$ appears again at $s$. This implies 
  \[ \frac{e}{c} = \sin(\gamma).\] 
  It is also easy to see that
  \[\sin(\gamma) = \frac{\dist(\ell(r_1,r_2),q)}{\dist(r_2,q)} \geq \frac{L^{-1}}{L} = L^{-2}.\]
  Further note that $\ell(r_2,q)$ must intersect $\seg(g,x)$ as $q$ blocks the visibility of $g$ but not of $x$. Thus $e=\dist(g,\ell(r_2,q))\leq \alpha$.
  This implies \[c = \frac{e}{\sin(\gamma)} \leq \alpha L^2.\]
  And thus it holds \[d\leq 2cL^{2} \leq 2 (\alpha L^2) L^2 = \alpha L^5 \leq L^{-7} L^5 = L^{-2}. \qedhere\]
\end{proof}

\begin{restatable}[Small Triangle]{lem}{SmallTriangle}\label{lem:SmallTriangleVisible}
   Let $x\in \P$ and let $r_1$ and $r_2$  be two vertices such that $\Delta = \Delta(x,r_1,r_2) \subseteq \P$. 
   Then it holds that $\Delta(x,r_1,r_2)$  is seen by $\grid^*(x)$, with $\alpha\leq L^{-7}$.
\end{restatable}

  \begin{figure}[htbp]
    \centering 
	\includegraphics{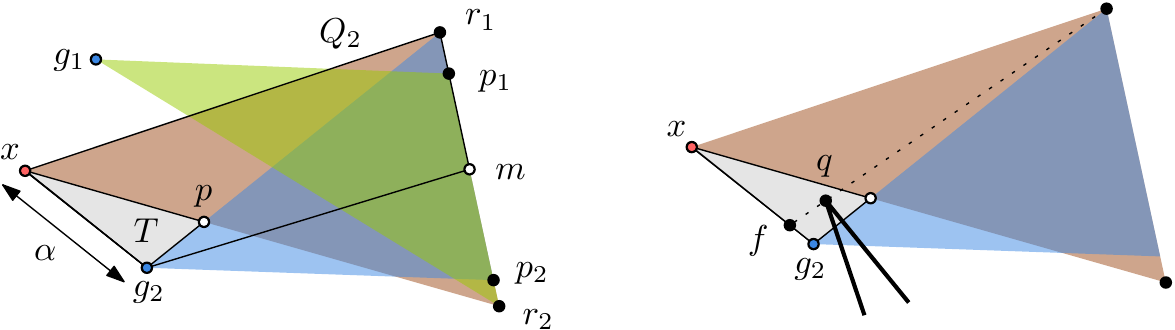}
      \caption{Illustration to the proof of Lemma~\ref{lem:SmallTriangleVisible}.}
	\label{fig:SmallTriangleVisible}
  \end{figure}

\begin{proof}
  We consider first the case that $\dist(x,\ell(r_1,r_2))\leq L^{-1}/2$.
  Note that there is at least one $g\in\grid(x)$ on the same side of
  $\ell(r_1,r_2)$ as $x$.
  It follows by Lemma~\ref{lem:distances} that there can be no vertex of
  \P blocking the visibility of $g$ to $\Delta(x,r_1,r_2)$.
  Thus from now on, we can assume that $\dist(x,\ell(r_1,r_2))\leq L^{-1}/2$ and this implies in particular that $x$ and $\grid(x)$ 
  are on the same side of $\ell(r_1,r_2)$.
  
  At first consider the case that there is a grid point $g\in\grid(x)$
  inside $\Delta$. 
  This implies the claim immediately, as $\Delta$ is convex and 
  it must be empty of obstructions, as $\Delta$ can be seen by $x$.

  So, let us assume that there exist two other grid points $g_1,g_2 \in \grid(x)$ as in Figure~\ref{fig:SmallTriangleVisible}.
  Let us first consider the case that there is no other vertex
  $q$ with $\dist(x,q)\leq L^{-1}$.
  This implies, we can use Lemma~\ref{lem:LimitedBlocking} and infer that
  the visibility regions of $g_1,g_2$ are only slightly blocked,
  as indicated in Figure~\ref{fig:SmallTriangleVisible}. In particular this implies that there exists a point $m\in \seg(r_1,r_2)$ that is visible from $g_1$ and $g_2$, as $\dist(r_1,r_2)\geq 1 \geq 2L^{-2}$.
  We define the quadrangle $Q_1 = Q(g_1,m,r_2,x)$ and $Q_2 = Q(g_2,m,r_1,x)$. Here $Q(t_1,t_2,t_3,t_4)$ indicates the quadrangle with vertices $t_1,\ldots,t_4$.
  Clearly it holds $\Delta\subseteq Q_1 \cup Q_2$.
  Thus it is sufficient to show that $g_i$ sees $Q_i$, for $i=1,2$.
  We define the point $p = \seg(g_2,r_1) \cap \seg(x,r_2)$ and the triangle 
  $T = \Delta(x,g_2,p)$. Inside $T$ is the only region where part of \P could block the visibility of $g_2$ to see $Q_2$ fully.
  As $g_2$ can be assumed to see $x$ this blocking part would correspond to a hole of \P. But note that a whole has at least $3$ vertices and area at least $1/2$. The area of $T$ is bounded by $\alpha L$, as one of its edges has length $\alpha$ and the height is trivailly bounded by $L$.
  
  It remains to consider the case that there exists one vertex $q$ with $\dist(x,q)<L^{-1}$. 
  This immediately implies $q\in \grid^*(x)$. 
  If $q$ does not block the vision of either $g_1$ or $g_2$,
  we are done.
  Otherwise, note that $q$ can block the vision of at most one of them, say $g_1$ and there is \emph{at most one} vertex $q$ with $\dist(x,q)< L^{-1}$.
  Thus after removal of $q$ the previous
  argument above can be applied.
  Now as $q$ blocks $g_1$ it must be either the case that $q\in \cone(g_1)$ 
  or $q\in T$.
  In the first case $q$ sees the part of $\cone(g_1)$ that is inside \P.
  In the second case, we denote by $f$ the point $\seg(g_1,x)\cap \ell(r_1,q)$. As $\dist(x,f)\leq \dist(x,g_2)$, we can conclude that 
  $f$ and $g_1$ see $\Delta$ by the same argument as above applied to $g_1$ and $f$ instead of $g_1$ and $g_2$, under the assumption that $q$ would not block $f$.
  Now however holds that $q\in \cone(f)$ and thus sees what $f$ would have seen and we are done.
\end{proof}

\begin{figure}[htbp]
    \centering 
	\includegraphics{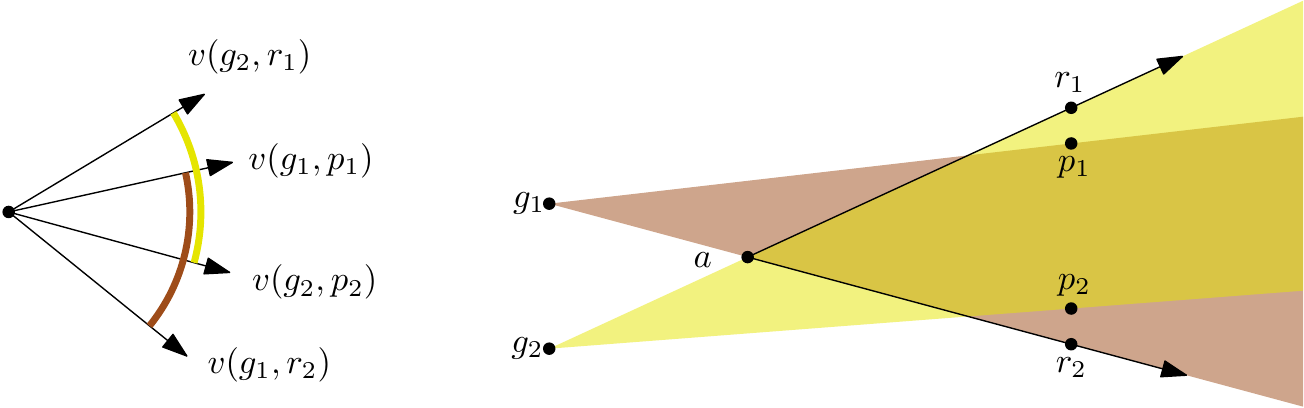}
      \caption{ Illustration to Definition~\ref{def:ConeProperty}.}
	\label{fig:ConePropertyRayimplCone}
\end{figure}

\begin{definition}[Cone-Property] \label{def:ConeProperty}
Given points $g_1,g_2$, two reflex vertices $r_1$ and $r_2$
and two points $p_1,p_2 \in \seg(r_1,r_2)$ with $\dist(p_1,r_1) \leq L^{-2} $ and $\dist(p_2,r_2) \leq L^{-2}$, see Figure~\ref{fig:ConePropertyRayimplCone}.
We denote by $C_1$ the cone with apex $g_1$ bounded by the rays $\ray(g_1,r_2)$ and $\ray(g_1,p_1)$
and we denote by $C_2$ the cone with apex $g_2$ bounded by the rays $\ray(g_2,r_1)$ and $\ray(g_2,p_2)$.
We say two points $g_1,g_2$ have the \emph{cone-property}
with respect to two reflex vertices $r_1,r_2$ if there exists some 
ray contained in $C_1\cap C_2$.

We define $a = \apex(g_1,g_2) = \seg(g_1,r_2)\cap \seg(g_2,r_1)$.
\end{definition}

The definition of $C_1$ and $C_2$ might seem a little odd,
but in spirit of Lemma~\ref{lem:LimitedBlocking}, 
we see that $C_i \subseteq \cone(g_i)$, for $i=1,2$,
if the conditions of Lemma~\ref{lem:LimitedBlocking} are met.

\begin{lemma}[New Cone]\label{lem:WholeCone}
  Let $g_1,g_2$ have the cone-property and assume the notation of Definition~\ref{def:ConeProperty}. Then the cone $C$ with apex $a$ 
  bounded by the rays $\ray(a,r_1)$ and $\ray(a,r_2)$ is contained 
  in $C_1\cup C_2$.
\end{lemma}
\begin{proof}
  The directions $v$ with $\ray(g_i,v) \subseteq C_i$, denoted by $I_i$, form an interval in $S^1$ for $i= 1,2$. 
  We define $I$ as the set of directions $v$ such that $ \ray(a,v) \subseteq C$.
  It is easy to see that $I = I_1\cup I_2$, because the cone-property implies that $I_1\cap I_2 \neq \varnothing$.
  Note that for all $b\in C_i$ holds $\ray(b,v) \subseteq C_i$, for all $i=1,2$ and $v\in I_i$.
  The fact $a\in C_1\cap C_2$ implies $C\subseteq C_1\cup C_2$.
\end{proof}

\begin{restatable}[Cone-Property]{lem}{ConeProperty}\label{lem:ConeProperty}
  Given two points $g_1,g_2$ with $\dist(g_1,g_2) \leq d \leq \tfrac{s}{4} \leq \tfrac{1}{4}$ 
  such that $g_1$ and $g_2$ are outside of 
  the embiggened $s$-bad region of the reflex vertices $r_1$ and $r_2$.
  Then $g_1$ and $g_2$ have the cone-property.
\end{restatable}

\begin{figure}[htbp]
    \centering

      \begin{minipage}[b]{.5\linewidth}
    \centering      \includegraphics{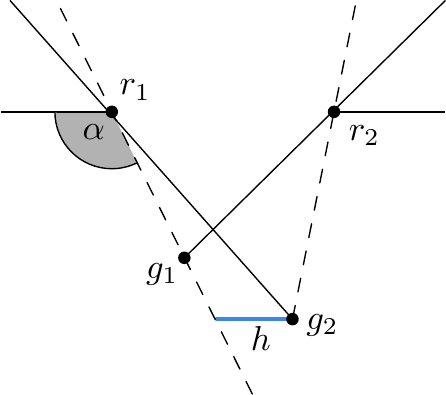}
    \subcaption{Definition of $\alpha$ and $h$.}\label{fig:Cones}
    \end{minipage}%
    \begin{minipage}[b]{.5\linewidth}
      \centering \includegraphics{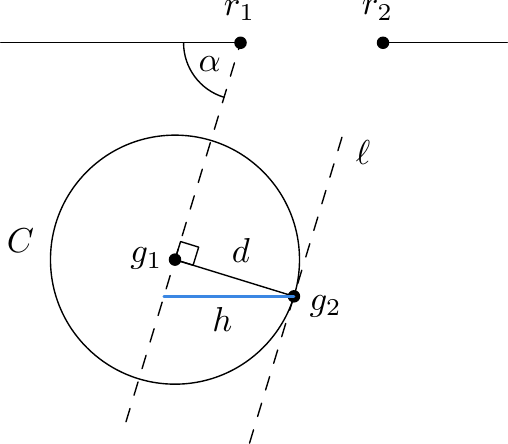}
      \subcaption{Definition of $\ell$ and $C$.}\label{fig:HorizontalDistance}
     \end{minipage}

     \noindent
     \begin{minipage}[b]{.49\linewidth}
      \centering \includegraphics{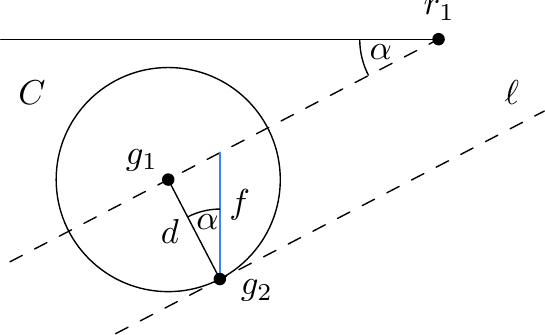}
      \subcaption{Illustrations for the case $\alpha < 45^{\circ}$}\label{fig:VerticalDistancePoints}
     \end{minipage}  
     \begin{minipage}[b]{.49\linewidth}
      \centering \includegraphics{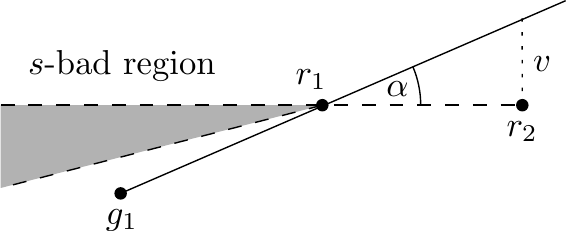}
      \subcaption{Illustrations for the case $\alpha < 45^{\circ}$}\label{fig:VerticalDistanceReflex}
     \end{minipage}  
  \caption{Illustrations to the proof of Lemma~\ref{lem:ConeProperty}}
    \label{fig:nolabel}
  \end{figure}

\begin{proof}
    It is sufficient to show that the rays $\ray(g_1,p_1)$ and $\ray(g_2,p_2)$ will not intersect, this is they are either parallel or diverge from one another.
    
    Note that $\dist(p_1,p_2) > \dist(r_1,r_2) - \dist(r_1,p_1) - \dist(r_2,p_2) \geq 1 - L^{-2} - L^{-2} > 1/2$. 
    Here, we used the fact that all vertices have integer coordinates.
    If we move $r_1$ towards $p_1$ and $r_2$ towards $p_2$ then 
    this does not change whether $\ray(g_1,p_1)$ and $\ray(g_2,p_2)$ will  intersect or not.
    Now, the assumption that $g_1$ and $g_2$ are not contained in the \emph{embiggened} $s$-bad region, becomes just that 
    $g_1$ and $g_2$ are not contained in the (ordinary) $s$-bad region.

    Thus from now on, we assume $r_1 = p_1$ and $r_2 = p_2$.
    We cannot make the assumption anymore that $r_1$ and $r_2$ have integer coordinates, but we can assume that $\dist(r_1,r_2) \geq 1/2$, which is sufficient for the rest of the proof.

    Without loss of generality, we assume that the supporting line of the two reflex vertices is horizontal. We can assume that the distance 
    $\dist(g_1,g_2) = d$, as this is the worst case.
    Further, we assume that $g_1$ is closer to $r_1$ than to $r_2$.

  For this purpose, we distinguish two different cases. Either the angle $\alpha$ between $\ell(r_1,r_2)$ and the ray $\ray(g_1,r_1)$  is  $\geq 45^{\circ}$  or  $< 45^{\circ}$ degree, see Figure~\ref{fig:Cones}.

  In the first case, we compare the \emph{horizontal} distance between the lines at two different locations. To be precise, we will show \[\dist_{\text{horizontal}}(\ell(g_1,r_1),g_2) < \dist_{\text{horizontal}}(\ell(g_1,r_1),r_2).\]

  This shows that the rays are in fact diverging. The horizontal distance between $r_2$ and $\ell(g_1,r_1)$ equals $\dist(r_1,r_2) \geq 1/2$ as was remarked above.
  The horizontal distance $h$ between the $\ell(g_1,r_1)$ and $g_2$ can be upper bounded by $\sqrt{2}d < 2 d$ as follows. See Figure~\ref{fig:HorizontalDistance} for an illustration.
  At first let $\overline{\ell}$ be a line parallel to $\ell(g_1,r_1)$ containing $g_2$. Then any point $p$ on $\overline{\ell}$ has the same horizontal distance to $\ell(g_1,r_1)$. Further $g_2$ has distance $d$ to $g_1$ and thus lies on the circle $C = \partial\disk(g_1,d)$ indicated in Figure~\ref{fig:HorizontalDistance}. The line $\ell$ parallel to $\ell(g_1,r_1)$ and furthest away from it that is still intersecting $C$ is indicated in Figure~\ref{fig:HorizontalDistance}. We can assume that $g_2$ lies on the intersection of $C$ and $\ell$, as $h$ would be smaller in any other case. We draw the horizontal segment $t$ realizing the horizontal distance between $g_2$ and $\ell(g_1,r_2)$. Note that the angle between $\ell(g_1,r_2)$ and $t$ equals $\alpha$. It is easy to see that $\seg(g_1,g_2)$ and $\ell$ are orthogonal. This implies that $\seg(g_1,g_2)$ and $\ell(g_1,r_1)$ are orthogonal as well.
  It follows 
  \[\sin \alpha = \frac{d}{h} \quad \Rightarrow \quad h = \frac{d}{\sin \alpha} \leq \sqrt{2}d < 2 d. \]
  Here we used the fact that $\sin \alpha \geq \tfrac{1}{
  \sqrt{2}}$, for $\alpha \geq 45^\circ$.
  In summary we have
  \[\dist_{\text{horizontal}}(\ell(g_1,r_1),g_2) \leq 2d \leq s/2\leq  1/2 \leq  \dist_{\text{horizontal}}(\ell(g_1,r_1),r_2).\]

  In the second case, we compare the \emph{vertical} distance $v$ between $\ell(g_1,r_1)$ and $r_2$ and $\ell(g_2,r_2)$ and $g_1$.
  We will show
  \[\dist_{\text{vertical}}(\ell(g_1,r_1),g_2) \leq \dist_{\text{vertical}}(\ell(g_1,r_1),r_2).\]
  By the same argument as in case one, we can conclude that 
  \[f =\dist_{\text{vertical}}(\ell(g_1,r_1),g_2)\leq \sqrt{2} d\leq 2 d.\]
  We repeat the argument to avoid potential confusion, see Figure~\ref{fig:VerticalDistancePoints}. 
  Let $\overline{\ell}$ be a line parallel to $\ell(g_1,r_1)$.
  Then every point $p$ on $\overline{\ell}$ has the same vertical 
  distance to $\ell(g_1,r_1)$. 
  Further $g_2$ has distance $d$ to $g_1$ and thus lies on the circle $C = \partial\disk(g_1,d)$ indicated in Figure~\ref{fig:VerticalDistancePoints}. The line $\ell$ parallel to $\ell(g_1,r_1)$ and furthest away from it that is still intersecting $C$ is indicated in Figure~\ref{fig:VerticalDistancePoints}. 
  We can assume that $g_2$ lies on the intersection of $C$ and $\ell$, 
  as $f$ would be smaller in any other case. 
  We draw the vertical segment $t$ realizing the vertical distance between $g_2$ and $\ell(g_1,r_2)$. Note that the angle between $\ell$ and $t$ equals $\alpha$. It is easy to see that $\seg(g_1,g_2)$ and $\ell$ are orthogonal. This implies that $\seg(g_1,g_2)$ and $\ell(g_1,r_1)$ are orthogonal as well.
  It follows 
  \[\cos \alpha = \frac{d}{f} \quad \Rightarrow \quad f = \frac{d}{\cos \alpha} \leq \sqrt{2}d < 2 d. \]
  Here we used the fact that $\cos \alpha \geq \tfrac{1}{
  \sqrt{2}}$, for $\alpha < 45^\circ$.

  Now, we give some bounds on the vertical 
  distance $v$ between $\ell(g_1,r_1)$ and $r_2$.
  See Figure~\ref{fig:VerticalDistanceReflex}. 
  Note that $\tan \alpha \geq s$ by the assumption that $g_1$ is not contained in an $s$-bad region.
  This implies 
  \[s \leq \tan \alpha = \frac{v}{\dist(r_1,r_2)} \leq \frac{v}{1/2} \leq 2v. \]
  
  In summary we have
  \[\dist_{\text{vertical}}(\ell(g_1,r_1),g_2) \leq 2d \stackrel{(1)}{\leq} s/2 \leq  v = \dist_{\text{vertical}}(\ell(g_1,r_1),r_2). \]
  For $(1)$ we used the assumption of the lemma.
\end{proof}

\begin{definition}[Cone Containment and Cutting Cones]
\label{def:CuttingCones}
We say  $\cone(x)$ is contained in $\cone(y)$ \emph{behind} $r_1$ and $r_2$ if 
$ \cone(x) \cap \ell^+ \subseteq \cone(y)\cap \ell^+$,  where $\ell^+$ 
is the half-plane bounded by $\ell(r_1,r_2)$ and does not contain the points 
$x$ and $y$.
When $r_1$ and $r_2$ are clear from context, we just say $\cone(x)$ is contained in $\cone(y)$.
In the same fashion, we define $\cone(z) = \cone(x)\cup \cone(y)$ \emph{behind} $r_1$ and $r_2$.
We say some cone $C$ is \emph{cut} by a line segment $s$ if the 
line segment $s' = C\cap s$ is non-empty and contains neither end point of $s$. 
\end{definition}

It is easy to see that for any $\cone(x)$ either holds that 
there exists a point $g\in \grid(x)$ with $g\in \cone(x)$ or
there exists two points $g_1,g_2\in \grid(x)$ such that
$\cone(x)$ cuts $\seg(g_1,g_2)$, see Figure~\ref{fig:GridTriangle}.

\begin{figure}[htbp]
  \centering
   \begin{minipage}[b]{.45\linewidth}
      \centering 
      \includegraphics{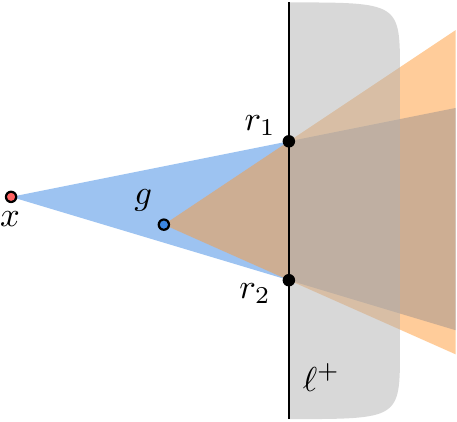}
      \subcaption{The point $g$ is contained in  $\cone(x)$ and thus 
	$\cone(g)$ contains $\cone(x)$ behind $\ell^+$.
	}
	\label{subfig:ConesContainment}
     \end{minipage} 
     \hspace{0.5cm}
     \begin{minipage}[b]{.45\linewidth}
      \centering 
      \includegraphics{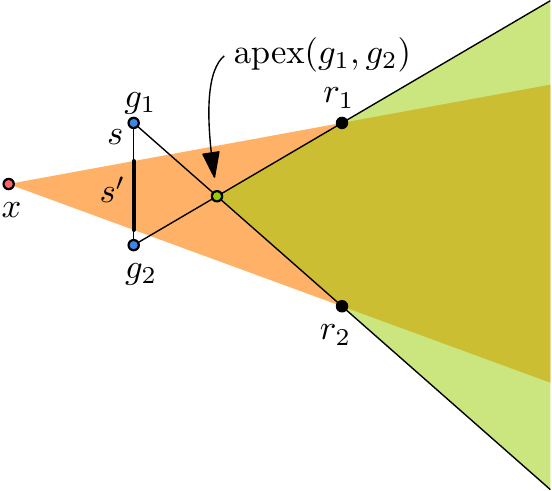}
      \subcaption{The line segment $s$ cuts $\cone(x)$, as $s'$ is non-empty and contains neither endpoint of $s$. Thus $\cone(x)$ contains $\apex(x,y)$.}
	\label{fig:ConeCutting}
     \end{minipage}

  \caption{Illustrations to cone containment and cone cutting.}
  \label{fig:ConesContainment}
\end{figure}

\begin{lemma}[Cone-Containment]\label{lem:ConesContainment}
  Consider the cones of $x$ and $g$ with respect to $r_1$ and $r_2$.
 Let $g \in \cone(x)$ then $\cone(x)$ is contained $\cone(g)$ behind $r_1$ and $r_2$.
\end{lemma}

\begin{lemma}[Cut-Segments]\label{lem:CutSegments}
  Consider the cones of $g_1,g_2$ and $x$ with respect to $r_1$ and $r_2$.
  Let $g_1,g_2$ have the cone-property and assume that $\cone(x)$ is cut by 
  $\seg(g_1,g_2)$. Then $\cone(x)$ is contained in $\cone(a)$, where $a = \apex(g_1,g_2)$ behind $r_1$ and $r_2$.
  Further $\cone(a) = \cone(g_1)\cup \cone(g_2)$ behind $r_1$ and $r_2$.
\end{lemma}

\begin{proof}
  We will show that $a \in \cone(x)$. 
  See Figure~\ref{fig:ConeCutting} for an illustration of the proof.
  Let $h_1$ be the half-plane bounded by $\ell(x,r_1)$ 
  containing $r_2$ 
  and let 
  $h_2$ be the half-plane bounded by $\ell(x,r_2)$
  containing $r_1$. 
  Recall that $a = \apex(g_1,g_2) = \seg(g_2,r_1)\cap \seg(g_1,r_2)$.
  Then it is clear that 
  $a \in \seg(g_2,r_1) \subseteq h_1$ and 
  $a \in \seg(g_1,r_2) \subseteq h_2$.
  Thus $a \in h_1 \cap h_2 = \cone(x)$.
\end{proof}

\begin{restatable}[Grid Outside Bad Region]{lem}{GridOutsideBad}\label{lem:OutsideBad}
  Let $x$ be a point not in the $s$-bad region of $r_1,r_2$, seeing $r_1$ and $r_2$, 
  and $\dist(x,r_i) \geq L^{-1}$, for $i=1,2$.
  Further we assume $ s \leq 8L\alpha $ and $L^{-3} \leq s $.
  Then it holds that  $\grid(x)$ is not in the embiggened $\tfrac{s}{2}$-bad region of $r_1$ and $r_2$.
\end{restatable}
 \begin{figure}[htbp]
      \centering
      \includegraphics{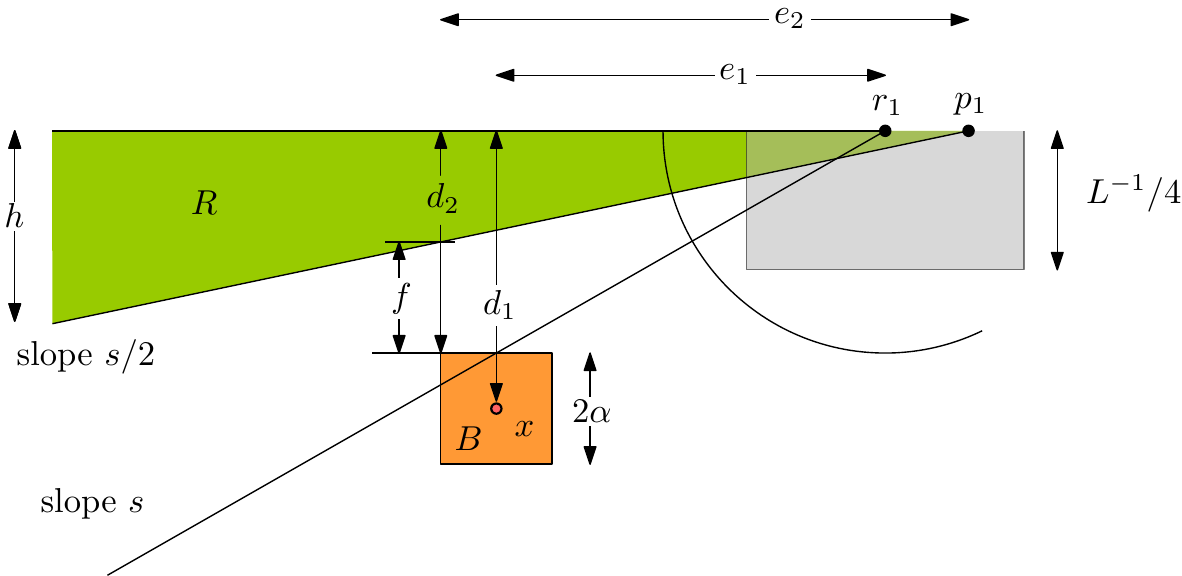}
      \caption{Illustrations to Lemma~\ref{lem:OutsideBad}.
	$\ell = \ell(r_1,r_2)$ is horizontal; 
	$B$ is an axis parallel box around $x$ with side length $2\alpha$;
	$R$ is an embiggened $(s/2)$-bad region;
	$e_1$ is the horizontal distance between $x$ and $r_1$; 
	$e_2 = e_1 + \alpha + L^{-2}$; 
	$d_1$ is the vertical distance between $\ell$ and $x$;  
	$d_2$ is the vertical distance between $\ell$ and $B$;
	$f = d_2 - (s/2)e_2$.
	}
      \label{fig:OutsideBad}
\end{figure}
\begin{proof}
  Refer to Figure~\ref{fig:OutsideBad} for 
  an illustration of this proof and the notation therein.
  We assume that $\ell(r_1,r_2)$ is horizontal
  and $x$ is closer to $r_1$ then to $r_2$.
  As every point of  $\grid(x)$ 
  has distance at most $\alpha$ from $x$, 
  it is sufficient to show that the box $B$ with sidelength 
  $2\alpha$ centered at $x$ does not intersect the embiggened $\tfrac{s}{2}$-bad 
  region $R$. 
  For the proof it is sufficient to assume that $x$ is outside the
  box $C$ of sidelength $L^{-1}/2$ centered at $r_1$.
  
  We denote by $h$ the largest width of $R$. It is easy to
  see that $h \leq (s/2) L \leq L^{-2}$.
  
  In case that $e_1 < L^{-1}/2$ holds $d_2 = d_1 - \alpha > L^{-1}/4 \gg h$. Thus, we assume from now on $e_1 \geq L^{-1}/2$.
  In this case, it holds
  \[e_2 = e_1 + L^{-2} + \alpha < \tfrac{3e_1}{2} .\]
  It suffices to show $f>0$.
\[
   f = d_2 - \frac{e_2  s}{2} 
   \stackrel{(1)}{>} d_2 - \frac{3 e_1 s}{4}  
   \stackrel{(2)}{=} d_1 - \alpha - \frac{3 e_1 s}{4}  
   \stackrel{(3)}{\geq} e_1 s - \alpha -  \frac{3 e_1 s }{4} 
   =  \tfrac{e_1 s}{4} - \alpha 
   \stackrel{(4)}{\geq}  \frac{L^{-1} s}{8} - \alpha 
   \stackrel{(5)}{>} 0.
\]


  Inequality $(1)$ was just shown above. 
  Equality $(2)$ can be easily seen in Figure~\ref{fig:OutsideBad}. 
  Inequality $(3)$ applies the fact that the $s$-bad region
  does not contain $x$ and has slope $s$. 
  Inequality $(4)$ follows the assumed lower bound  $e_1 > L^{-1}/4$.
  Inequality $(5)$ is the main assumption of the lemma.
\end{proof}
We are now ready to proof Lemma~\ref{lem:SpecialLocalVisibility}.

\SpecialVisa*

 \begin{figure}[htbp]
    \centering 
	\includegraphics{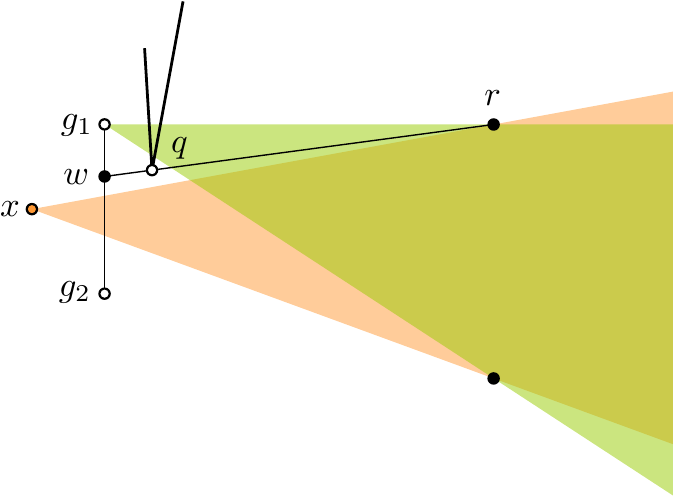}
      \caption{In this scenario there exists a vertex $q$, which blocks the visibility
      $g_1$. Luckily, we can define another point $w$, which is not blocked by $q$. And $q$ can see the cone of $w$.}
	\label{fig:HandleCloseBlocking}
\end{figure}

\begin{proof}[Proof of Lemma~\ref{lem:SpecialLocalVisibility}]
   Note first that the triangle $\Delta(x,r_1,r_2)$ is seen by $\grid^*(x)$
 by Lemma (Small Triangle)~\ref{lem:SmallTriangleVisible}.
 So, we are only interested in $\cone(x)$ behind $r_1$ and $r_2$.

The easiest case to be ruled out is that there exists $g\in\grid^*(x)$
 such that $g\in \cone(x)$, as by Lemma (Cone-Containment)~\ref{lem:ConesContainment} 
 this implies the claim. 
 So from now on, we assume for all $g \in \grid^*(x)$ holds $g\notin \cone(x)$.

 We know that there exists two grid points $g_1$ and $g_2$ 
 such that $\cone(x)$ cuts the segment $\seg(g_1,g_2)$.
 By Lemma (Cut-Segments)~\ref{lem:CutSegments} it remains to show that 
 $g_1$ and $g_2$ have the Cone-Property.
 For this purpose, we want to invoke Lemma (Cone-Property)~\ref{lem:ConeProperty}.
 To this end, we have to invoke a series of other lemmas.
 
 Note that $r_1,r_2\in \cone(x)$ 
 and   $\dist(x,r_i) < L^{-1}$, for $i=1$ or $i=2$ implies $r_i$ is included in  $\grid^*(x)$.
 Thus by the argument above, we assume from now on $\dist(x,r_i)\geq L^{-1}$ for all $i=1,2$. 
 There might still be a different vertex $q\neq r_1,r_2$ with $\dist(x,q)< L^{-1}$.
 We deal first with the case that 
 there are no vertices $q$ with $\dist(x,q)< L^{-1}$.
 By Lemma (Grid Outside Bad Region)~\ref{lem:OutsideBad} holds that  $\grid(x)$ is not contained in the embiggened $s'$-bad region with $s' = s/2$ with respect to $r_1$ and $r_2$. 
 We can apply Lemma (Limited Blocking)~\ref{lem:LimitedBlocking} as $\alpha<L^{-7}$ and what is said above.
 From Lemma (Limited Blocking)~\ref{lem:LimitedBlocking} follows that $\cone(g_i)$ contains $C_i$, for $i=1,2$ as defined in Definition (Cone-Property)~\ref{def:ConeProperty}, see also Definition (Cone-Property)~\ref{def:VisibilityCones} to recall the definition of $\cone(g_i)$. 
 
 It remains to show that $g_1$ and $g_2$ satisfy the Cone-Property.
 To this end, we need to show that the assumptions of Lemma (Cone-Property)~\ref{lem:ConeProperty} is met.
 Here, we consider the embiggened $s'$-bad region with $s' = s/2$.
 Note that \[\dist(g_1,g_2) \leq 2\alpha \leq s/(8L) < s'/4  = s/8 <1/4.\]
 This shows the claim together with Lemma (New Cone)~\ref{lem:WholeCone}.

 It remains to consider the case that there exists one vertex $q$ with $\dist(x,q)<L^{-1}$, see Figure~\ref{fig:HandleCloseBlocking}. 
 This immediately implies $q\in \grid^*(x)$. 
 If $q$ does not block the vision of either $g_1$ or $g_2$,
 we are done.
 Otherwise, note that $q$ can block the vision of at most one of them, say $g_1$ and there is \emph{at most one} vertex $q$ with $\dist(x,q)< L^{-1}$.
 We define $w = \ell(q,r)\cap \seg(g_1,g_2)$. As the edges incident to $q$ block $g_1$ at least partially, we know that $w$ exists.
 As $\dist(w,g_2)\leq \dist(g_1,g_2)$ and $\cone(x)$ cuts the segment $\seg(w,g_2)$, we can use the same arguments as above.
 By definition of $w$ holds $q\in \cone(w)$.
 And behind $r_1,r_2$ the $\cone(q)$ contains $\cone(w)$. 
\end{proof}

\subsection{Global Visibility Containment}
This simple lemma quantifies (as a function of $s$ and $L$) the maximum \emph{width} of $s$-bad regions.

\begin{lemma}[distance bad region to supporting line]\label{lem:BRSL}
Let $p$ be a point of \P inside an $s$-bad region associated to opposite reflex vertices $r_1$ and $r_2$.
Then $\dist(p,\ell(r_1,r_2)) \leq sL$.
\end{lemma}

\begin{proof}
See Figure~\ref{fig:NormalBadRegion37}.
\end{proof}

Although it is not possible to achieve a local visibility containment property for all points in \P, the exceptions only involve bad regions. 
Under Assumption~\ref{ass:generalPos} (and Assumption~\ref{ass:integer}), we can give a fairly short proof of Lemma~\ref{lem:GridReplacement}.
As preparation, we need the following technical lemma which heavily relies on Lemma~\ref{lem:distances}.
  
  \begin{figure}[htbp]
  \centering
    \includegraphics{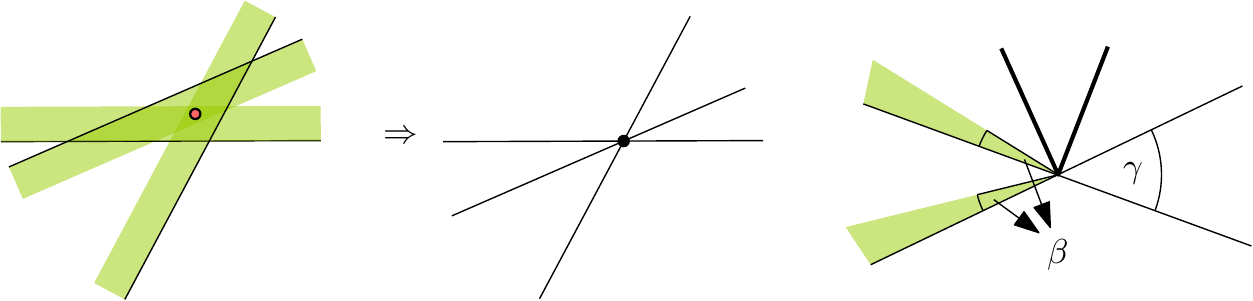}
  \caption{Three bad regions meeting in an interior point implies that the extensions must meet in a single point. No two bad regions intersect in the vicinity of a vertex, as they are defined by some angle $\beta \ll L^{-2}$.
  But the angle $\gamma$ between any two extensions is at least $L^{-2}$.}
  \label{fig:NoThreeBad}
\end{figure}
  
\begin{lemma}[no three bad regions intersect]\label{lem:NoBadIntersection}
  Under Assumption~\ref{ass:integer} and~\ref{ass:generalPos}, for any  $s \leq L^{-9}$, no point in the interior of \P belongs to three different $s$-bad regions.
\end{lemma}
\begin{proof}
We consider now the case that there exists a point $x$ with $\dist(x,v)\leq L^{-2}$, for some vertex $v$. We show that $x$ is contained in at most \emph{one} bad regions, see to the right of Figure~\ref{fig:NoThreeBad}.
note first that any extension $\ell$ with $v\notin \ell$ has distance at least $L^{-1}$ from $v$, by Lemma~\ref{lem:distances}. And thus $x$ cannot be in any bad region belonging to $\ell$ by Lemma~\ref{lem:BRSL}.
By Lemma~\ref{lem:distances} the angle between any two extensions must be at least $L^{-2}$. As we have at most two vertices contained on any line, the bad regions belonging to the extensions through $v$ must start at $v$, see Definition~\ref{def:BadRegion}.
For the angle $\beta$ defining the bad regions holds $\tan(\beta) = s\ll L^{-2}$
and thus all bad regions in the vicinity of $v$ are disjoint.
(Note that $v$ itself is not considered as part of the bad region.)

Let $\ell_1,\ell_2,\ell_3$ be supporting lines of three distinct pairs of opposite reflex vertices.
As we assumed that no three points lie on a line, those three lines are also distinct.
We first consider the case where two of those supporting lines, say $\ell_1$ and $\ell_2$ are parallel. 
By Lemma~\ref{lem:distances} Item~\ref{itm:parralelLines}, $\dist(\ell_1,\ell_2) \geq L^{-1}$.
Also, by Lemma~\ref{lem:BRSL}, any point of an $s$-bad region is at distance at most $sL$ of the corresponding supporting line.
Therefore, any point in the intersection of the $s$-bad region associated to $\ell_1$ and the one associated to $\ell_2$ is at distance at most $L^{-8}$ from those two lines; a contradiction to $\dist(\ell_1,\ell_2) \geq L^{-1}$.

We now show that any intersection of two supporting lines (among $\ell_1,\ell_2,\ell_3$) should be in the interior of \P.
Such an intersection cannot be on the boundary of \P deprived of the vertices of \P, since it would immediately yield three supporting lines meeting in a point.
If two supporting lines, say $\ell_1$ and $\ell_2$, meet in a vertex of \P, then this vertex is one of the opposite reflex vertices for both $\ell_1$ and $\ell_2$ (otherwise there would be three vertices on a line).

Assume now that the intersection $p$ of say, $\ell_1$ and $\ell_2$ is outside \P.
By Lemma~\ref{lem:distances} Item~\ref{itm:IntersectionIntersection}, the distance of $p$ to any point in \P is at least $L^{-5}$.
Let $p'$ be a point of \P in the intersection of two $s$-bad regions associated to $\ell_1$ and to $\ell_2$.
By Lemma~\ref{lem:BRSL}, the distance of $p'$ to both $\ell_1$ and $\ell_2$ is at most $L^{-8}$.
That implies, by setting $d$ to $L^{-8}$ in Lemma~\ref{lem:distances} Item~\ref{itm:LinesCommonIntersection}, that $\dist(p',p) \leq L^{-8}L^2=L^{-6}$; a contradiction to $\dist(p,\P) \geq L^{-5}$  

Thus, we can suppose that $\ell_1,\ell_2,\ell_3$ pairwise intersect in three distinct points $p=\ell_1 \cap \ell_2,q=\ell_1 \cap \ell_3,r=\ell_2 \cap \ell_3$ in the interior of \P; this is because we assumed that no three extensions meet in a point.
Let $p'$ be in the $s$-bad regions associated to $\ell_1$ and $\ell_2$.
As explained in the end of the previous paragraph, $\dist(p',p) \leq L^{-6}$.
By Lemma~\ref{lem:distances} Item~\ref{itm:IntersectionLine}, $\dist(p,\ell_3) \geq L^{-5}$.
By Lemma~\ref{lem:BRSL} any point in the $s$-bad region associated to $\ell_3$ is at distance at most $L^{-8}$.
As $L^{-6}+L^{-8} < L^{-5}$, $p'$ can not be in the $s$-bad region associated to $\ell_3$.
Which means that the intersection of the three $s$-bad regions associated to $\ell_1$, $\ell_2$, and $\ell_3$ is empty.   
\end{proof}

   \begin{figure}[htbp]
  \centering
    \includegraphics{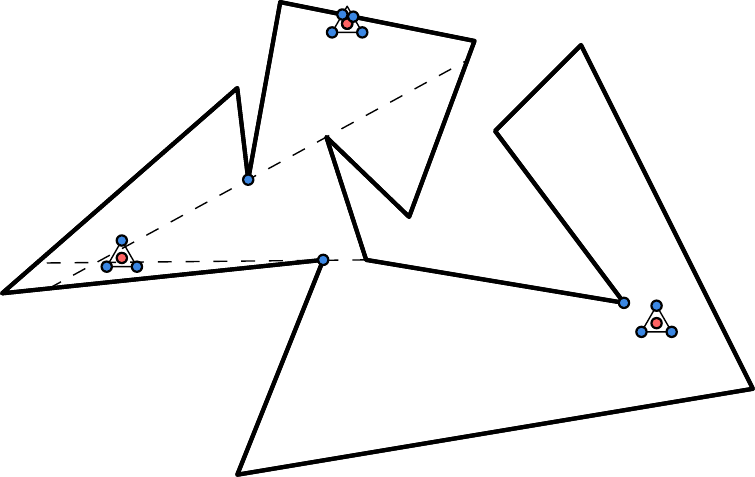}
  \caption{The red dots indicate the optimal solution. The blue dots indicate the  
  The red dot on the top is in the interior case and four grid points are added around it.
  The red dot on the left is too close to two supporting lines and we add one of the reflex vertices of each of the supporting lines. 
  The red dot to the right has distance less than $L^{-1}$ to a reflex vertex, so we add that vertex as well to $G$.}
  \label{fig:GlobalVisbility}
\end{figure}
  
\begin{proof}[Proof Lemma~\ref{lem:GridReplacement} using Assumptions \ref{ass:integer} and \ref{ass:generalPos}.]
  We denote by $OPT$ an optimal solution of size $k$.
  We assume that no point of $OPT$ is actually contained
  in $\Gamma$ as we can just take that point into our grid solution.
  In particular this implies $OPT$ contains non of the vertices of \P.
  Let $\alpha = L^{-11}$ and  $s<L^{-9}$.
  Let $x\in \P$ be some point and $R(x)$ some set of size at most $2$ that contains a reflex vertex for each $s$-bad region, that $x$ is contained in.
  As no point is contained in three bad regions $R(x)$ having size $2$ is enough, see Lemma~\ref{lem:NoBadIntersection}.
  We define \[G = \bigcup_{x\in OPT}\grid^*(x) \, \cup\, R(x).\]
  It is easy to see that $G\subseteq \Gamma$ has size $|G| \leq (7+2)k$.  We want to argue that $G$ sees the entire polygon.
  For each $x\in OPT$ the local containment property holds, except for the bad regions it is in, see Lemma~\ref{lem:SpecialLocalVisibility}. These parts are seen by the reflex vertices we added.  
\end{proof}

\section{Conclusion}\label{sec:conclusion}

We presented an $O(\log |OPT|)$-approximation algorithm for the \pgag problem under two relatively mild assumptions.
The most natural open question is whether Assumption~\ref{ass:generalPos} can be removed.
We believe that this is possible but it will require some additional efforts and ideas.
Another improvement of the result would be to achieve an approximation ratio of $O(\log n)$ for polygons with holes.
This would match the currently best known algorithm for the \textsc{Vertex 
Guard} variant and known lower bounds.
In that respect, it might be very useful that Lemma~\ref{lem:GridReplacement} does not require the polygon to be simple.
One might also ask about the inapproximability of \pgag for simple polygons.
For the moment, the problem is only known to be inapproximable for a certain constant ratio (quite close to 1), unless P=NP.
It would be interesting to get superconstant inapproximability under standard complexity theoretic assumptions or improved approximation algorithms.

\newpage
\bibliographystyle{abbrv}
\bibliography{art-gallery}

\end{document}